\PassOptionsToPackage{table}{xcolor}
\PassOptionsToPackage{hyphens}{url}

\documentclass[letterpaper,twocolumn,10pt]{article}
\usepackage{usenix2019_v3}

\usepackage{cite}
\usepackage{amsmath,amssymb,amsfonts}
\usepackage{algorithmic}
\usepackage{graphicx}
\usepackage{textcomp}
\usepackage{xcolor}
\usepackage{lscape}
\usepackage{tcolorbox}
\usepackage{graphics}
\usepackage{booktabs}
\usepackage{amsmath,amssymb,amsfonts}
\usepackage{algorithmic}
\usepackage{graphicx}
\usepackage{textcomp}
\usepackage{xcolor}
\usepackage{wasysym}
\usepackage{multirow}
\usepackage{makecell}
\usepackage{subcaption}
\usepackage{enumitem}
\usepackage{listings}
\usepackage{soul,color}
\usepackage{subcaption}
\usepackage{hyperref}
\usepackage{soul}
\usepackage[framemethod=TikZ]{mdframed}
\usepackage{wrapfig}
\usepackage{flushend}
\usepackage{tablefootnote}
\usepackage[numbers]{natbib}
\usepackage[affil-it]{authblk}

\makeatletter
\def\@fnsymbol#1{\ensuremath{\ifcase#1\or \dagger\or \ddagger\or
   \mathsection\or \mathparagraph\or \|\or **\or \dagger\dagger
   \or \ddagger\ddagger \else\@ctrerr\fi}}
\makeatother

\def\BibTeX{{\rm B\kern-.05em{\sc i\kern-.025em b}\kern-.08em
    T\kern-.1667em\lower.7ex\hbox{E}\kern-.125emX}}

\hypersetup{
    linkcolor=blue,
}

\newcommand*\circled[1]{\tikz[baseline=(char.base)]{
            \node[shape=circle,draw,inner sep=0.5pt] (char) {#1};}}

\title{\textsc{A New Hope}: Contextual Privacy Policies for Mobile Applications and An Approach Toward Automated Generation}

\author[1,2\thanks{The author completed most of the research while being a visiting student at Singapore Management University}]{Shidong Pan}
\author[1,2]{Zhen Tao}
\author[2\thanks{Corresponding Author}]{Thong Hoang}
\author[1,2]{Dawen Zhang}
\author[3]{Tianshi Li}
\author[1,2]{Zhenchang Xing}
\author[2]{Xiwei Xu}
\author[2]{Mark Staples}
\author[2]{Thierry Rakotoarivelo}
\author[4]{David Lo}
\affil[1]{\textit{School of Computing, Australian National University}}
\affil[2]{\textit{Software and Computational Systems Research Program, CSIRO's Data61}}
\affil[3]{\textit{Khoury College of Computer Sciences, Northeastern University}}
\affil[4]{\textit{School of Computing and Information
Systems, Singapore Management University}}

\date{}

\begin{document}
\maketitle

\pagestyle{plain}

\begin{abstract}
Privacy policies have emerged as the predominant approach to conveying privacy notices to mobile application users. In an effort to enhance both readability and user engagement, the concept of contextual privacy policies (CPPs) has been proposed by researchers.
The aim of CPPs is to fragment privacy policies into concise snippets, displaying them only within the corresponding contexts within the application's graphical user interfaces (GUIs). 
In this paper, we first formulate CPP in mobile application scenario, and then present a novel multimodal framework, named \textbf{\textsc{SeePrivacy}}, specifically designed to automatically generate CPPs for mobile applications. 
This method uniquely integrates vision-based GUI understanding with privacy policy analysis, achieving 0.88 precision and 0.90 recall to detect contexts, as well as 0.98 precision and 0.96 recall in extracting corresponding policy segments.
A human evaluation shows that 77\% of the extracted privacy policy segments were perceived as well-aligned with the detected contexts. 
These findings suggest that \textsc{SeePrivacy} could serve as a significant tool for bolstering user interaction with, and understanding of, privacy policies. Furthermore, our solution has the potential to make privacy notices more accessible and inclusive, thus appealing to a broader demographic.
A demonstration of our work can be accessed at  \url{https://cpp4app.github.io/SeePrivacy/}

\end{abstract}



\section{Introduction}~\label{sec_intro}
Privacy policies have become the most prevalent privacy notice approach for mobile applications, with the aim of protecting users' privacy and digital security~\cite{flavian2006consumer, perez2018review, harkous2018polisis, kemp2020concealed, adams2020agreeing, harkous2022hark, bui2023detection, li2023s}.
On the contrary, application users often find themselves in a privacy paradox, specifically, \textit{Digital Resignation}. This term refers to a state where users desire to understand and control the information that digital entities own about them, but feel powerless to do so~\cite{draper2019corporate}.
A study has found that the average length of privacy policies is around 4,000 words for 75 prominent mobile applications and websites, requiring an estimated 16 minutes to read for an average adult~\cite{Blakkarly2022privacy}. Consequently, according to a survey conducted by Obar and Oeldorf-Hirsch~\cite{obar2020biggest}, a significant portion of users, approximately 74\%, opt for the ``quick join'' clickwrap option without engaging with privacy policies, primarily due to the lengthy and detailed nature of these documents.
These numbers indicate that privacy policies have become increasingly difficult to read and understand in a timely manner, exacerbating the prevalence of digital resignation among users.

\begin{figure} [t!]
\begin{subfigure}{.324\linewidth}
  \centering
  \includegraphics[width=.98\linewidth]{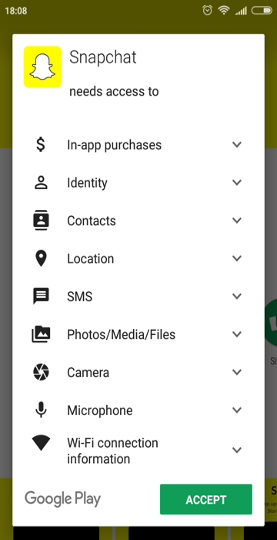}
  \caption[]{Install-time}
  \label{fig_intro_install}
\end{subfigure}%
\hspace{3.15pt}
\begin{subfigure}{.315\linewidth}
  \centering
  \includegraphics[width=.98\linewidth]{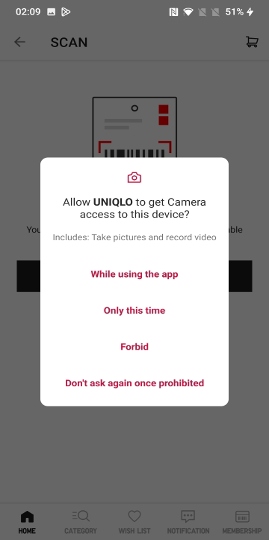}
  \caption{Invoke-time}
  \label{fig_intro_invoke}
\end{subfigure}
\hfill
\begin{subfigure}{.315\linewidth}
  \centering
  \includegraphics[width=.967\linewidth]{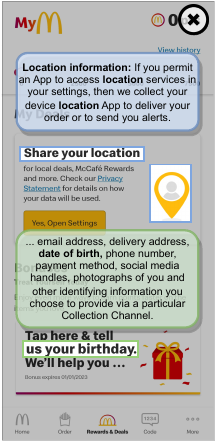}
  \caption{Context-aware}
  \label{fig_intro_context}
\end{subfigure}
\caption[Caption]{
Figure~\ref{fig_intro_install} represents the install-time reminder of required permissions~\cite{intro_installtime}; Figure~\ref{fig_intro_invoke} corresponds to the invoke-time reminder that appears when permission is first invoked; Figure~\ref{fig_intro_context} presents our proposed contextual privacy policy, specifically designed for mobile applications.}
\label{fig_introduction}
\vspace{-5pt}
\end{figure}

In addressing the pressing issue of digital resignation, the contemporary evolution of privacy policies has given rise to the concept of ``just-in-time'' privacy notices~\cite{ICOjustintime}. These notices furnish users with detailed information about data collection at the point of interaction. This practice has been recognized and included in the U.K. General Data Protection Regulation as a recommended approach for crafting privacy policies~\cite{ICOjustintime}. However, the dominant application of ``just-in-time'' notices is primarily limited to the installation phase of applications and first invocation of certain permissions. 
Figure~\ref{fig_intro_install} shows the ``install-time'' permission reminders that appear when a user installs a new Android app\footnote{This feature was essentially deprecated after Android 6.0 (API level 23) since developers must request permissions at runtime.}, delineating the permissions to be granted. 
Figure~\ref{fig_intro_invoke} presents another example, which illustrates a camera access reminder triggered at the initial ``invoke-time''. 
However, there are inherent limitations to current ``just-in-time'' notices. For one, while they provide a snapshot of when an app accesses specific data, it doesn't necessarily offer insights into the broader context -- how this data might subsequently be transmitted off the device, shared with third parties, stored, or used for other purposes. 
Furthermore, current implementations of ``just-in-time'' reminders fall short in terms of efficacy, as security warnings regarding device permissions tend to be isolated from their immediate context, thus failing to adequately engage user awareness~\cite{felt2010effectiveness, ma2014android}.

To counteract these deficiency, researchers proposed \textit{Contextual Privacy Policies} (CPPs)~\cite{Feth2017, windl2022automating}.
This innovative concept aims to deconstruct traditional privacy notices into shorter and digestible segments, displaying them only in relevant contexts in mobile applications.
Privacy-related contexts refer to areas in which data practices are directly associated with graphical user interfaces (GUIs). These contexts on the screen are identified and the corresponding privacy policy segments are retrieved and displayed, with key information highlighted.
This integration of privacy notices within the GUIs ensures a close and timely alignment with the context, thus conveying precise and succinct privacy information.
\citet{windl2022automating} demonstrated that CPPs have the potential to address the usability issues in privacy policies, thereby enhancing user engagement and awareness of data practices.
They also presented a framework for automatically generating CPPs for websites.
Studies have shown that users spend 90\% of their time using mobile apps rather than websites~\cite{Mobiloud, Amplitude2022, AppVelocity2022}. 
However, the prior~\cite{windl2022automating} framework has the following limitations when applied to the mobile app context, all of which are successfully addressed in this work:
1) Their context selection is based on empirical observation of users' behaviour in website browsing scenarios, but lacks connection to privacy regulations and existing standards in practice.
2) The prior context detection method is based on ad-hoc HTML analysis, which has limited internal generalizability, especially for more compact and complicated user interfaces in mobile app platforms.
Overall, while the concept of CPP has a great potential, generating accurate CPPs for mobile apps proposes unique challenges that cannot be addressed by the previous automated generation framework.

%

In this paper, we first formulate the CPPs in the mobile scenario, and then propose a novel multimodal framework, dubbed \textbf{\textsc{SeePrivacy}}, designed to automatically generate CPPs for mobile applications.
Figure~\ref{fig_intro_context} presents an example of our CPP in mobile applications. 
Within this framework, computer vision techniques
are incorporated with pre-trained large language models (LLMs) to analyze mobile GUI screenshots, thus pinpointing privacy-related contexts. In addition, natural language processing techniques
are used to dissect privacy policy texts, extracting the corresponding policy sections for identified contexts.
To quantitatively evaluate performance, we establish our benchmark dataset, named \textbf{\textsc{Cpp4App}}, which covers more than 1,200 privacy-related contexts along with their matching privacy policy segments in mobile settings.
The quantitative evaluation demonstrates the robust performance of our framework in multiple evaluation metrics. 
Regarding the identification of privacy-related contexts, \textsc{SeePrivacy} achieves accuracy, precision, and recall values of 0.81, 0.88, and 0.90, respectively. 
Regarding the extraction of the corresponding segments, the framework attains accuracy, precision, and recall rates of 0.94, 0.98, and 0.96, respectively.
To further assess the functionality of \textsc{SeePrivacy}, we performed a human evaluation with 15 examiners on about 120 CPPs. The results are encouraging: 77\% of the extracted privacy policy segments were perceived as elucidating the detected context. 


We also discussed the applications, implications, and
contributions of CPPs and \textsc{SeePrivacy} to the broader community. In the survey, participants display a significantly increased willingness to engage with CPPs, achieving an average score of 4.1 out of 5. In contrast, traditional privacy policies received an average score of just 2 out of 5. 
This difference suggests \textsc{SeePrivacy} could contribute meaningfully to improve user engagement with privacy policies, potentially mitigating the widespread challenge of digital resignation.
Additionally, the overall usefulness rating for \textsc{SeePrivacy} is 4.53 out of 5, indicating strong user endorsement. 
Furthermore, we elaborate on two potential application scenarios of \textsc{SeePrivacy}: displaying CPPs in app markets and detecting ``lack of disclosure'' with exact problematic context.
Lastly, we discuss how CPPs potentially make the privacy policies more inclusive and allow it to reach a larger audience such as children.

In summary, our proposed framework extends the automated generation of CPPs to mobile applications, which are the digital platforms users spend the majority of time on.
The key contributions are:
\begin{itemize} [leftmargin=*, noitemsep, topsep=0pt]
\item To our knowledge, this is the first research to define the concept of contextual privacy policies (CPPs) within mobile application scenarios. 
\item We have introduced a novel framework, named \textbf{\textsc{SeePrivacy}}, designed to automatically generate these contextual privacy policies for mobile applications.  
\item We have constructed the benchmark dataset, named \textbf{\textsc{Cpp4App}}, specifically to evaluate contextual privacy policies on mobile platforms and to quantitatively assess the functionality of the \textsc{SeePrivacy} framework. 
\end{itemize}

\noindent \textbf{Roadmap.} The remainder of this paper is structured as follows: Section~\ref{sec_background} introduces the development of transparent privacy notices and privacy-related context identification; 
Section~\ref{sec_dataset} introduces our benchmark dataset, \textsc{Cpp4App}; Section~\ref{sec_framework} details the design of our proposed framework, i.e.,\textsc{SeePrivacy}; and Section~\ref{sec_evaluation} presents the results of evaluating \textsc{SeePrivacy} on the benchmark dataset; Section~\ref{sec_discussion} discusses the implications and contributions of our work to the broader community; and Section~\ref{sec_conclusion} concludes the paper.

\noindent \textbf{Ethical Considerations and Responsible Disclosure.} Ethical approval for this research was secured from our institution's Institutional Review Board (IRB). In conducting our user study, we avoided collecting any personally identifiable information (PII). In line with our commitment to responsible research practices, any privacy discrepancies detected during the course of the study were promptly and transparently disclosed to both the study participants and the corresponding application developers.

\section{Background}
\label{sec_background}

%
\begin{table*}[!t]
  \caption{Tallies of data types' existence.}
  \label{tab_survey}
  \centering
  \resizebox{.95\textwidth}{!}{%
  \begin{tabular}{l|cccccc|cccccc}
    \hline
    \rowcolor{lightgray!15}
    &                                                                                        \multicolumn{6}{c|}{Basic Personal Identifiable Information}                                                                       & \multicolumn{6}{c}{Other Personal Information}                                                                                     \\ 
    \hline
    \rowcolor{lightgray!50}
    &\textbf{Name} & \textbf{Birthday} & \textbf{Address} & \textbf{Phone} & \textbf{Email} &\textbf{Profile}
    & \textbf{Contacts} & \textbf{Location} & \textbf{Photos} & \textbf{Voices} & \textbf{Financial Info} 
    & \textbf{Social Media} \\
    \hline
    
    \rowcolor{lightgray!85}
    \textbf{Administration}
    & & & & & & & & & & & &\\
    \rowcolor{lightgray!15}
    GDPR~\cite{GDPR} 
    & \CIRCLE &\Circle &\CIRCLE &\CIRCLE &\CIRCLE &\CIRCLE
    & \Circle &\Circle &\Circle &\Circle &\Circle &\Circle 
    \\
    \rowcolor{lightgray!50}
    CCPA~\cite{CCPA} 
    & \CIRCLE &\Circle &\Circle &\Circle &\CIRCLE &\CIRCLE
    & \Circle &\CIRCLE &\Circle &\Circle &\CIRCLE &\Circle 
     \\
    \rowcolor{lightgray!15}
    CalOPPA~\cite{CalOPPA} 
    & \CIRCLE &\Circle &\CIRCLE &\CIRCLE &\CIRCLE &\Circle
    & \Circle &\CIRCLE &\Circle &\Circle &\CIRCLE&\CIRCLE 
    \\
    \rowcolor{lightgray!50}
    COPPA~\cite{COPPA} 
    & \CIRCLE &\Circle &\CIRCLE &\CIRCLE &\CIRCLE &\CIRCLE
    & \Circle &\CIRCLE &\CIRCLE &\CIRCLE &\Circle&\Circle 
    \\
    
    \rowcolor{lightgray!15}
    APP~\cite{APPs} 
    &\CIRCLE &\Circle &\CIRCLE &\CIRCLE &\CIRCLE &\CIRCLE
    &\Circle &\CIRCLE &\CIRCLE &\Circle &\CIRCLE &\Circle
    \\
    
    \rowcolor{lightgray!85}
    \textbf{Industry}
    & & & & & &  & & & & & & \\
    \rowcolor{lightgray!15}
    Google Play~\cite{googledatasafety} 
    & \CIRCLE & \CIRCLE & \CIRCLE & \CIRCLE & \CIRCLE & \CIRCLE
    & \CIRCLE & \CIRCLE & \CIRCLE & \CIRCLE & \CIRCLE & \CIRCLE
    \\
    \rowcolor{lightgray!50}
    Apple App~\cite{appleprivacy} 
    & \CIRCLE & \Circle & \CIRCLE & \CIRCLE & \CIRCLE & \CIRCLE
    & \CIRCLE & \CIRCLE & \CIRCLE & \CIRCLE & \CIRCLE & \CIRCLE
    \\
    \rowcolor{lightgray!15}
    Huawei AppGallery~\cite{Huaweiprivacy} 
    & \CIRCLE & \CIRCLE & \CIRCLE & \CIRCLE & \CIRCLE & \CIRCLE
    & \CIRCLE & \CIRCLE & \CIRCLE & \CIRCLE & \CIRCLE & \CIRCLE 
    \\
    \rowcolor{lightgray!50}
    Amazon Appstore~\cite{Amazonprivacy} 
    & \Circle & \Circle & \Circle & \CIRCLE & \CIRCLE & \Circle
    & \CIRCLE & \CIRCLE & \CIRCLE & \CIRCLE & \CIRCLE & \CIRCLE
    \\
    \rowcolor{lightgray!15}
    Samsung Galaxy Store~\cite{samsungprivacy}
    & \Circle & \Circle & \Circle & \CIRCLE & \Circle & \Circle 
    & \CIRCLE & \CIRCLE & \CIRCLE & \CIRCLE & \Circle & \Circle
    \\

  \hline
\end{tabular}
}%
\end{table*}

Mobile applications provide users with unparalleled convenience, multiple entertainment options, and various services. At the same time, they are responsible for the collection and exchange of substantial amounts of personal information, raising concerns surrounding data privacy. In response to the potential risks of privacy violations, the U.S. Department of Health, Education, and Welfare proposed a set of Fair Information Practice Principles (FIPP) in 1973~\cite{FIPP1973, landesberg1998privacy}. Later, the Federal Trade Commission adopted these principles as ``\textit{the notice and choice privacy framework}''~\cite{obar2020biggest}, which is commonly regarded as the foundation for modern privacy regulations such as the European General Data Protection Regulation (GDPR)~\cite{GDPR}, the California Consumer Privacy Act (CCPA)~\cite{CCPA}, and the Australian Privacy Principles (APP)~\cite{APPs}.
In FIPP, the most fundamental principle is \textit{``notice''}~\cite{FTC_FIPP}. Consumers should be provided with a clear understanding of an entity's information practices before any personal data is collected from them (e.g., [GDPR Art. 12, GDPR Rectical 58]). In the absence of transparent and understandable notice, it becomes impossible for a consumer to make an informed decision regarding the extent of personal information they choose to disclose.


\subsection{Related Work}
Research efforts to improve the presentation and readability of privacy policies have been manifold and encompass various approaches. Kelly et al.~\cite{kelley2009nutrition, kelley2010standardizing, kelley2013privacy} introduced an innovative method called privacy nutrition labels (PNLs), designed to facilitate consumers' understanding of how their information is collected and utilized in a structured table-like way. By employing this concise and organized framework, PNLs effectively aid in understanding the privacy practices of software applications~\cite{ cranor2012necessary, li2022understanding, li2022understanding123, pan2023toward}. In a related vein, Cranor~\cite{cranor2002web} proposed a protocol, i.e., Platform for Privacy Preferences, that allows website organizations to declare their intended use of information collected from website users. Alongside these approaches, there has been a growing focus on visual aids such as privacy icons~\cite{holtz2011towards, holtz2011privacy, efroni2019privacy}. These small graphical symbols, symbolizing privacy concepts or practices, serve as visual shorthand, simplifying complex privacy policies. By standardizing these icons across various services and platforms, researchers aim to reduce confusion, thereby making privacy policies more readily accessible to the general populace~\cite{holtz2011privacy}.

The conceptualization of \textit{Contextual Privacy Policies} emerged in 2004, marking a significant change in the approach to the presentation of privacy notices. Bolchini et al.~\cite{bolchini2004need} discerned that the conventional reading of privacy policies represented a monolithic block that hindered users from retrieving essential privacy details. In response, they invented a systematic method aimed at reorganizing privacy policies around specific user interaction contexts. Subsequent to this development, Feth~\cite{Feth2017} extended the concept by introducing \textit{contextual privacy statements}. 
Rather than using a ``one-size-fits-all'' privacy policy that must suit every usage scenario, these contextual privacy statements offer precise privacy and data protection information more closely with the user's immediate context.
In a recent advancement, Windl et al.~\cite{windl2022automating} proposed the design and architecture of PrivacyInjector, a production AI tool that can automatically generate contextual privacy policies for websites.
However, their scope of data practice identification is limited to six website-specific types, such as “Advertising” and “Cookies and Tracking Elements”. Their identification procedure primarily considers which date type users are likely to encounter when browsing websites. In contrast, our approach is regulation-driven, examining which data practices should be transparently disclosed in privacy notices.
Moreover, their context detection method is intrusive and based on ad-hoc HTML analysis, which is consequently not applicable to mobile applications. 
In contrast, the source code of mobile applications is typically inaccessible to end-users. This limitation led us to develop a purely vision-based context detection method, tailored to the unique challenges presented by mobile applications.
Nonetheless, CPPs have been empirically shown to increase privacy notice transparency, constituting a progressive step toward more user-centric privacy communication
~\cite{ortloff2020implementation, masotina2022transparency}.



Contrasting with websites, mobile applications represent a unique landscape for personal information collection, often involving multifaceted approaches that include device sensors, user input, and third-party integrations~\cite{pan2023large}. These complex processes necessitate a precise and transparent articulation of how user information is acquired, employed, and disseminated—an issue of critical concern. Given the expansive and continually growing scale of mobile applications, there exists an emergent need for an automated framework capable of generating CPPs tailored specifically to this domain. In pursuit of this objective, the present work attempts to establish such a framework. As a preliminary step, we recognize the imperative to define clear and comprehensive definitions for privacy-related contexts within the mobile environment, forming the foundation for our proposed framework.

\subsection{Empirical Data Practice Identification}
\label{sec_background_survey}

According to Windl et al.~\cite{windl2022automating}, privacy-related contexts are considered areas on graphical user interfaces (GUIs) where data practices might be applied.
To rigorously define these practices, we formulate two guiding questions, each tailored to address a distinct facets of the issue:

\noindent \textit{\textbf{1.}} What data practices should be included in privacy notices by regulations?

\noindent \textit{\textbf{2.}} What data practices are of concern and desire for users?

%
These questions collectively serve to demarcate the scope of our investigation, guiding our analysis in a manner that encompasses regulatory compliance and user-centric considerations.
%
To our knowledge, there are no existing studies that specifically address the issue of contextual privacy policies within the domain of mobile applications. Specifically, current research appears to be primarily concerned with improving the presentation and readability of privacy policies~\cite{kelley2009nutrition, kelley2013privacy, kelley2010standardizing, cranor2002web, ciocchetti2008future}. Therefore, we perform a rigorous examination of data practices as they relate to privacy regulations and industry market standards (see Table~\ref{tab_survey}). We present each group as follows:



\noindent\textbf{\emph{Administration.}} The administration often functions as the key regulator within the broader ecosystem of data privacy. A report by the United Nations Conference on Trade and Development reveals that 137 of 194 countries and regions have instituted data protection and privacy legislation~\cite{UNCTAD}. To gain insight into the administrative perspective on these matters, we examine five representative privacy regulations, including the GDPR~\cite{GDPR}, CCPA~\cite{CCPA}, the California Online Privacy Protection Act (CalOPPA)~\cite{CalOPPA}, the Children's Online Privacy Protection Act (COPPA)~\cite{COPPA}, and APP~\cite{APPs}. These regulations serve as critical touchpoints in our exploration of the administration's stance on data practices.


\noindent \textbf{\emph{Industry.}} Mobile application platforms typically impose particular privacy notice requirements on curated applications within their marketplaces, with special attention to the categories of data involved. For instance, the Google Play app store has delineated 14 primary data categories and 39 sub-categories in their data safety reports, which all applications must adhere to~\cite{googledatasafety}. These standards have been formulated through rigorous user studies and market tests, reflecting a comprehensive understanding of user concerns and expectations. Such market standards provide valuable insight into the alignment of privacy requirements with actual user needs and industry practices, thereby serving as a meaningful reference for privacy policy development in the mobile application domain.

\subsection{Multimodal Privacy-related Contexts}
\label{sec_background_context}

\begin{table*}[!t]

\caption{Privacy-related contexts.}
  \label{tab_keyword_list}
\centering

\resizebox{1.0\textwidth}{!}{%
\begin{tabular}{|l|l|l|l|c|}
\hline
\textbf{Data types} & \textbf{Description} & \textbf{Privacy-related keywords lists for textual GUI elements}  & \textbf{Related RICO-icon class} & \textbf{Icon examples}\\ 
\hline
\hline
Name & How a user refers to themselves & \makecell[l]{name, first name, last name, full name, real name, surname,\\ family name, given name}  & n.a. &  
\raisebox{-0.0\totalheight}{%
n.a.
\begin{minipage}{.04\textwidth}
  \includegraphics[width=0.85\linewidth]{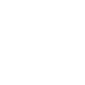}
\end{minipage}
}%
\\ \hline
Birthday & A user’s birthday & birthday, date of birth, birth date, DOB, birth year & n.a. & 
\raisebox{-0.0\totalheight}{%
n.a.
\begin{minipage}{.04\textwidth}
  \includegraphics[width=0.85\linewidth]{Figures/placeholder.png}
\end{minipage}
}%
\\ \hline
Address & A user’s address  & \makecell[l]{mailing address, physical address, postal address, billing address, \\shipping address, residential address, residence, personal address} &n.a& 
\raisebox{-0\totalheight}{%
n.a
\begin{minipage}{.04\textwidth}
  \includegraphics[width=0.85\linewidth]{Figures/placeholder.png}
\end{minipage}
}%
\\ \hline
Phone & A user’s phone number & \makecell[l]{phone, phone number, mobile phone, mobile number, telephone, \\call, telephone number}  & [43] Call &
\raisebox{-0.06\totalheight}{%
\begin{minipage}[c]{.04\textwidth}
  \includegraphics[width=0.92\linewidth]{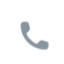}
\end{minipage}
}%
 \raisebox{-0.06\totalheight}{%
\begin{minipage}[c]{.04\textwidth}
  \includegraphics[width=0.8\linewidth]{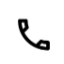} 
\end{minipage}
}%
\\ \hline
Email & A user’s email address  
 \raisebox{-0\totalheight}{%
\begin{minipage}{.04\textwidth}
  \includegraphics[width=0.85\linewidth]{Figures/placeholder.png}
\end{minipage}
}%
& email, e-mail, email address, e-mail address  & [6] Email & 
\raisebox{-0.06\totalheight}{%
\begin{minipage}[c]{.04\textwidth}
  \includegraphics[width=0.85\linewidth]{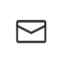}
\end{minipage}
}%
 \raisebox{-0.05\totalheight}{%
\begin{minipage}[c]{.04\textwidth}
  \includegraphics[width=0.8\linewidth]{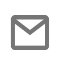} 
\end{minipage}
}%
\\ \hline
Profile & A user’s account information 
 \raisebox{-0\totalheight}{%
\begin{minipage}{.04\textwidth}
  \includegraphics[width=0.8\linewidth]{Figures/placeholder.png}
\end{minipage}
}%
& profile, account
& [49] Avatar & 
\raisebox{-0.06\totalheight}{%
\begin{minipage}[c]{.04\textwidth}
  \includegraphics[width=0.85\linewidth]{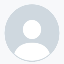}
\end{minipage}
}%
 \raisebox{-0.05\totalheight}{%
\begin{minipage}[c]{.04\textwidth}
  \includegraphics[width=0.8\linewidth]{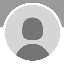} 
\end{minipage}
}%
\\ \hline
\hline
Contacts &  \makecell[l]{A user’s contact information, or the \\access to the contact permission} 
& contacts, phone-book, phone book, device's address book  &[68] Group, [3] Follow &
\raisebox{-0.05\totalheight}{%
\begin{minipage}[c]{.04\textwidth}
  \includegraphics[width=0.8\linewidth]{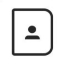}
\end{minipage}
}%
 \raisebox{-0.05\totalheight}{%
\begin{minipage}[c]{.04\textwidth}
  \includegraphics[width=0.8\linewidth]{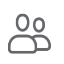} 
\end{minipage}
}%
\\ \hline
Location &  \makecell[l]{A user’s location information, or the \\access to the location permission} & \makecell[l]{location, locate, 
geography,\\ geo, geo-location, precision location} & \makecell[l]{[40] Location crosshair, \\  \text{[72] Location}} &
\raisebox{-0.03\totalheight}{%
\begin{minipage}[c]{.04\textwidth}
  \includegraphics[width=0.8\linewidth]{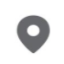}
\end{minipage}
}%
\raisebox{-0.05\totalheight}{%
\begin{minipage}[c]{.04\textwidth}
  \includegraphics[width=0.8\linewidth]{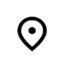}
\end{minipage}
}%
\\ \hline
Photos & \makecell[l]{A user’s photos, videos, or the access \\to the camera permission} & \makecell[l]{camera, photo, scan, album, picture, gallery, photo library,\\ storage, image, video, scanner, photograph} & \makecell[l]{[42] Photo, [56] Videocam,\\ \text{[82] Wallpaper}}  &
\raisebox{-0.05\totalheight}{%
\begin{minipage}[c]{.04\textwidth}
  \includegraphics[width=0.8\linewidth]{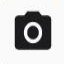}
\end{minipage}
}%
\raisebox{-0.05\totalheight}{%
\begin{minipage}[c]{.04\textwidth}
  \includegraphics[width=0.8\linewidth]{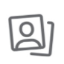}
\end{minipage}
}%
\\ \hline
Voices&  \makecell[l]{A user’s voices, recordings, or the access\\ to the microphone permission} & microphone, voice, mic, speech, talk, audio & [91] Microphone  & 
\raisebox{-0.05\totalheight}{%
\begin{minipage}[c]{.04\textwidth}
  \includegraphics[width=0.75\linewidth]{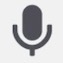}
\end{minipage}
}%
\raisebox{-0.05\totalheight}{%
\begin{minipage}[c]{.04\textwidth}
  \includegraphics[width=0.75\linewidth]{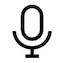}
\end{minipage}
}%
\\ \hline
Financial info & \makecell[l]{Information about a user’s financial \\accounts, purchases, or transactions}  & \makecell[l]{credit card, company, companies, organization, commercial,\\ organizations, pay, payment, financial, bill, wallet, purchase} & [61] Cart  &
\raisebox{-0.05\totalheight}{%
\begin{minipage}[c]{.04\textwidth}
  \includegraphics[width=0.85\linewidth]{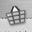}
\end{minipage}
}%
\raisebox{-0.05\totalheight}{%
\begin{minipage}[c]{.04\textwidth}
  \includegraphics[width=0.8\linewidth]{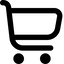}
\end{minipage}
}%
\\ \hline
Social media &  \makecell[l]{A user's social media information, or \\the access to social media accounts}
& social media, Facebook, Twitter, socialmedia, share  & [77] Facebook, [89] Twitter &
\raisebox{-0.05\totalheight}{%
\resizebox{.042\textwidth}{!}{%
\begin{minipage}[c]{.041\textwidth}
  \includegraphics[width=0.8\linewidth]{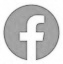}
\end{minipage}
}%
}%
\raisebox{-0.05\totalheight}{%
\resizebox{.04\textwidth}{!}{%
\begin{minipage}[c]{.04\textwidth}
  \includegraphics[width=0.8\linewidth]{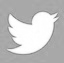}
\end{minipage}
}%
}%
 \\ \hline
\end{tabular}
}%
\end{table*}

Privacy-related contexts refer to areas within GUIs where data practices might be applied. In this investigation, we categorize two primary types of GUI components that indicatively collect personal information via human-device interactions: literal texts and graphical icons. Literal texts, often regarded as the foundational information sources in mobile application GUIs, are juxtaposed with graphical icons, which are ingeniously designed to communicate vital information with maximum efficiency. The growing adoption of graphical icons in place of literal texts is due to their concise nature, which provides users with the requisite information instantaneously, thereby substantially minimizing the cognitive effort required to interpret the accompanying textual descriptions~\cite{xiao2019iconintent}.


\noindent \textbf{\emph{Keyword collection.}} 
In the process of analyzing literal texts that describe data types, we snowballed a keyword list tailored to CPPs. 
This list was compiled by extracting keywords from academic research literature about privacy policies, including content categorization~\cite{wilson2016creation, kaur2018comprehensive, zimmeck2019maps, li2021developers}, automated generation~\cite{yu2015autoppg, sun2020quality, zimmeck2021privacyflash}, and compliance analysis~\cite{yu2018enhancing, zimmeck2019maps, zhang2020does, xie2022scrutinizing, andow2020actions, cui2023poligraph}.
These studies often present unique challenges due to differences in privacy policies, corpus selections, and specific tasks among various topics. 
Ambiguous terms, such as ``address,'' which could refer to a residential or an email address, were deliberately excluded to avoid confusion. To further promote unification and consistency across contexts, we rendered the list case-insensitive. Table~\ref{tab_keyword_list} presents a comprehensive display of the finalized keyword list of each data type.


\noindent \textbf{\emph{Icon collection.}}
Icons frequently serve as symbols for specific categories of privacy information. One common example is that the camera-like icons typically correspond to the authorization of camera access by mobile applications or access to the device photo gallery; and a location crosshair icon often represents the access to location permission or users physical location information. Liu et al.~\cite{Liu:2018:LDS:3242587.3242650} developed a comprehensive icon taxonomy, comprising 99 common icon classes in mobile applications. This research effort also yielded one of the largest icon datasets, consisting of over 118,000 distinct icons. Subsequently, we performed a semantic alignment between the RICO-icon classes proposed by Liu et al. and the data types that were identified through our survey.
By doing so, the textual information (privacy-related keywords) and visual information (icons) on GUI are well aligned with regulatory requirements (data types), as detailed in Table~\ref{tab_keyword_list}.
\section{\textsc{Cpp4App}: The Benchmark Dataset}
\label{sec_dataset}

In our study, the construction of a benchmark dataset is essential to quantitatively assess the efficacy of the proposed framework in generating contextual privacy policies (CPPs). Although existing datasets are available for privacy policy analysis~\cite{wilson2016creation, zimmeck2019maps, andow2020actions, amos2021privacy}, and for understanding mobile graphical user interfaces (GUIs)~\cite{deka2017rico, leiva2020understanding, bunian2021vins}, we identified two major shortcomings within current collections. These limitations, as detailed below, necessitate the compilation of a new dataset.
\begin{itemize} [leftmargin=*, noitemsep, topsep=0pt]

\item \textbf{Temporal inconsistency between applications, policies, and GUIs.} 

With mobile applications undergoing frequent changes in privacy practices, there is a concomitant need to update their corresponding privacy policies. In addition, the GUIs of these apps are subject to rapid evolution. Consequently, it becomes imperative to maintain alignment between the screenshots (representing the GUIs) and their temporally corresponding privacy policies for CPP generation. This alignment ensures the precision and dependability of quantitative evaluations.

\item \textbf{No annotations about privacy-related contexts in GUIs.} 

To the best of our knowledge, we are the first to introduce the concept of CPPs specific to mobile applications. For the purpose of evaluating the performance of the automated generation framework, benchmark datasets should ideally incorporate two essential features: 1) the annotated regions (i.e., bounding boxes) representing privacy-related contexts within GUI screenshots, and 2) the mapping relations that connect these privacy-related contexts to corresponding segments of the privacy policy. Unfortunately, as of the time of this study, no existing dataset encompasses both of these vital characteristics.
\end{itemize}
In the following subsections, we present steps employed to create a benchmark dataset, thereby addressing the aforementioned issues.

\subsection{Mobile Apps Selection}

The Android platform is often considered transparent and favorable for academic research, with numerous apps markets readily available~\cite{wei2017survey, li2019rebooting}. Among these, the Google Play app store stands out as the most expansive and accessible, hosting more than two million applications~\cite{malavolta2015end, wang2018beyond}. In particular, Google Play is a leader in requiring mobile applications to provide a link to their privacy policy on their homepages~\cite{googledeveloper}. As a result, users downloading Android mobile applications from the Google Play app store can directly access the respective privacy policy links. 

Consistent with previous research, widely used mobile applications tend to adhere to more stringent privacy practices~\cite{zimmeck2019maps, liu2021have, USENIX_2022_GEODIFF}. 
In addition, less popular applications often have poorly written privacy policies~\cite{pan2023large, pan2023toward}, making it meaningless to generate CPPs for them.
In this study, we leveraged market data from AppBrain~\cite{appbrain}, one of the largest third-party platforms monitoring app statistics for the Google Play app store, to pinpoint the 200 most popular mobile applications. 
Our study collection method includes four specific strategies. 
First, recognizing the variations in GUIs, functions, and privacy practices across different application categories (i.e., genre), we endeavored to ensure a diverse representation by selecting the most popular mobile applications from 17 distinct categories. 
Second, applications published by the same parent company normally share the same privacy policy, so we try to select apps from different developers for better representation.
Third, we excluded any mobile applications with privacy policies that were less than 200 words or smaller than 2 KB in size, as these may constitute invalid or insufficient privacy policies~\cite{USENIX_2022_GEODIFF, liu2023appcorp}. 
Finally, we limited our selection to those applications that offered privacy policies in English, thus narrowing our candidate pool to 50 mobile applications for subsequent fine-grained annotation.

\subsection{Context Annotation Strategy}


To create a benchmark dataset, it is imperative to ascertain that vital privacy-related GUI elements are incorporated into our dataset. Therefore, drawing on established guidelines about mobile application GUI design~\cite{10essentialscreens, 9mainscreens}, we have systematically gathered specific pages from each selected application.
The collection process adheres to the categorizations delineated below:
\begin{itemize}[leftmargin=*, noitemsep, topsep=0pt]
\itemsep0em
\item \emph{Registration/Login page.} 
Mobile applications frequently gather user registration information on this page to facilitate account creation. This may encompass both personal and other private details. Certain applications require new users to register before accessing specific features, making it infeasible for these users to circumvent the registration or login process.

\item \emph{Home page.} This serves as the initial interface upon accessing a mobile application. It contains essential features and information related to its app's functionality, which may necessitate the provision of personal data or the granting of device permissions. Furthermore, screenshots of adjacent main pages are also collected for analysis.

\item \emph{Profile page.} This section may retain users' personal information. For example, some mobile applications might request access to the camera or photo roll to furnish customization options related to public profile appearance.

\item \emph{Setting page.} Often associated with privacy customization, this page may govern permissions related to device sensors and access to personal information.

\item \emph{Onboarding page.} It serves as a tutorial or introduction. Moreover, this page elucidates a mobile application's salient features and benefits.

\item \emph{Map page.} Common to various application categories such as travel, posting, food delivery, and shopping, this page is often utilized to amass users' location data.

\item \emph{Essential pages for specific categories.} 
This includes specialized interfaces such as the product card pages in shopping apps, the post-editing interface in forum applications, and the feed page within social media platforms.
\end{itemize}
We follow these categorizations to systematically collect relevant pages from each selected mobile applications, ensuring that essential privacy-related GUI elements are represented in our dataset. 
Notably, we have opted for manual collection of GUI pages as opposed to relying on static analysis for several reasons. 
Accurate static analysis is a challenging task in mobile program analysis, particularly for popular apps that are not only complexly programmed but also commonly employ code obfuscation.
Furthermore, iterating through all possible activities to capture their corresponding GUIs is a non-trivial task, and outdated or restricted GUIs may not be accessible through regular user interactions. Manual collection allows us to naturally avoid the aforementioned problems.

After the data collection, the annotators need to manually examine the text in each screenshot, cross-referencing it with our CPP-related keyword list, as mentioned in Table~\ref{tab_keyword_list}. If the text contains any of these CPP-related keywords, it is labeled as the corresponding data type and designated with a bounding box. Subsequently, the annotators are tasked with detecting all icons in the screenshot to determine if any belong to a privacy-related data type. If an icon meets this criterion, it is labeled with the corresponding class and marked with a bounding box. For these purposes, the \textit{OpenCV-Python}~\cite{OpenCV} library is employed to draw bounding boxes and document coordinates. Two annotators were asked to execute the annotation process independently. For any disagreement, they discussed and agreed on the same opinion, and if the disagreement persisted, a senior annotator joined the discussion to facilitate a resolution.
The percent agreement (inter-rater reliability) for initial annotation is about 84\%.


\begin{table}[!t]
\centering
  \caption{Basic statistic of \textsc{Cpp4App}.}
  \label{tab_dataset_summary}
  \resizebox{0.8\linewidth}{!}{%
  \begin{tabular}{l | c}
    No. Mobile applications \& Privacy policies & 50  \\
    \hline
    \hline
    No. Screenshots & 402 \\
    Screenshots per Mobile application & 8.04\\
    No. Privacy-related contexts & 1,217\\
    Contexts per Mobile application & 24.34\\
    \hline
    \hline
    No. Words in Privacy policies & 297,010 \\
    Words per Privacy policy & 5,940\\
\end{tabular}
}%
\end{table}

\begin{figure*}[t]
  \centering
  \includegraphics[width=.8\linewidth]{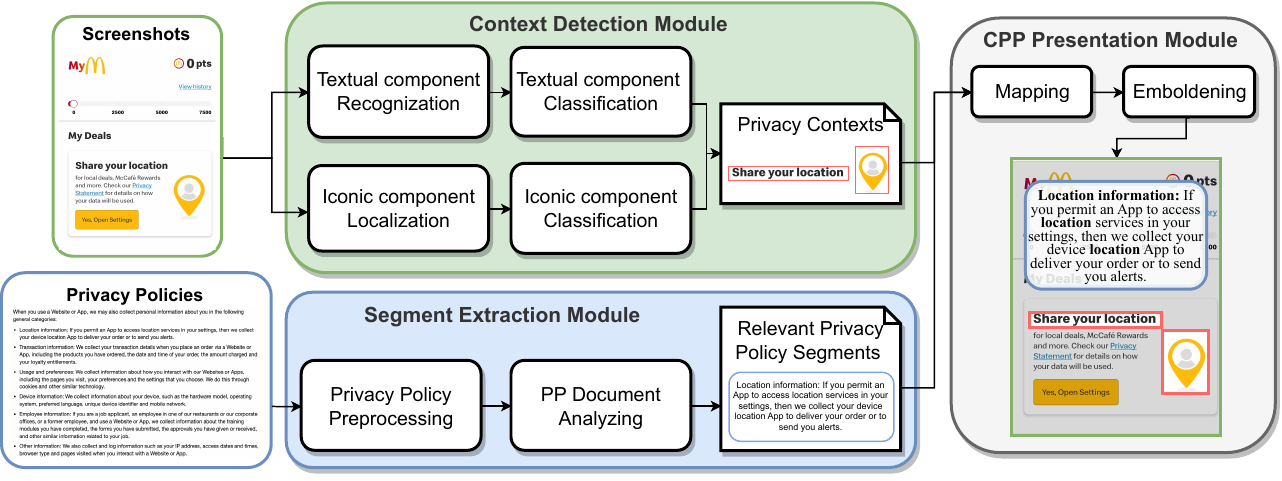}

  \caption{An overview of \textsc{SeePrivacy}. 
  }
  \label{fig_pipeline}
\end{figure*}

\subsection{Segment Annotation Strategy}

Initially, if annotators are unfamiliar with the mobile application of the policy they need to annotate, they are mandated to explore and familiarize themselves with its features for a duration of 10 minutes. 
Then, two annotators independently scrutinize each data privacy-related context within the privacy policies and extract pertinent segments. To ensure the integrity and quality of the annotation process, we establish a reading speed parameter ranging from 250 to 300 words per minute (the reading speed of an average adult~\cite{Blakkarly2022privacy}) for the annotators, implementing countdown timers in accordance with the policy lengths. 
If the $segment\_sim$ (See the Formula~\ref{formula_similarity}) is above 0.8, two annotations are counted as an agreement.
The mechanism for resolving disagreements is consistent with the process outlined in the previous paragraph. 
In cases where both annotators are unable to identify any relevant segments within the privacy policies, the segment is explicitly labeled as ``No relative information is found in the privacy policy.''

\subsection{Setups and Statistics}
\label{dataset_setup}

All annotators involved in this study are research scientists with a minimum of two years of experience in privacy research for mobile applications. To ensure the alignment between the GUIs and privacy policies, an Android mobile phone, i.e., the OnePlus 7 Pro, Android 11, and Hydrogen OS 11.0.9.1.GM21, is employed to download all the selected mobile applications. The corresponding privacy policies are downloaded simultaneously. To preserve the status quo of the apps and policies, the auto-update features within the Google Play app store, individual mobile apps, and the system's general settings on the phone (including auto-update over Wi-Fi) are all deactivated. In total, this effort resulted in the collection of 402 screenshots from 50 diverse mobile apps, accompanied by 1,217 labeled privacy-related contexts, each correlating to specific segments within privacy policies. We have designated this benchmark dataset as \textsc{Cpp4App}, and the statistical analysis is provided in Table~\ref{tab_dataset_summary}. As the inaugural dataset of its nature, we are confident that it is sufficient for evaluating our automatic generation framework and positioned to serve as a benchmark for future research in this field.

\section{\textsc{SeePrivacy}: The Framework}~\label{sec_framework}
In this section, we introduce our multimodal framework \textsc{SeePrivacy}, designed to automatically generate CPPs for a given screenshot and a given privacy policy, as illustrated in Figure~\ref{fig_pipeline}. The framework accepts privacy policies and screenshots from a given mobile application as input parameters. These inputs are processed through three primary modules within the framework to autonomously build CPPs for the designated mobile application. The three main modules integral to our framework are detailed as follows:

\begin{itemize}[leftmargin=*, noitemsep, topsep=0pt]
\itemsep0em
\item \textit{Context Detection Module:} This module is designed to identify specific contexts within a mobile application's screenshots where CPPs should be displayed. 
\item \textit{Segment Extraction Module:} This module focuses on extracting pertinent segments from the privacy policy of the given mobile application.
\item \textit{CPPs Presentation Module:} This module is responsible for linking privacy policy segments to their corresponding contexts and articulating the essential information, thereby facilitating the generation of CPPs for mobile applications. 
\end{itemize}
The details of each module are explained in the following subsections. 

\subsection{Context Detection Module}~\label{sec_framework_cdm}
The source code of mobile apps is typically inaccessible to end-users. 
This limitation led us to develop a purely vision-based Context Detection Module tailored to the unique challenges presented by mobile apps.
Specifically, in our approach to GUI understanding, we opt for visual components as atomic elements rather than function-level GUI widgets. This choice is driven by the observation that certain widgets (e.g., a navigation bar) may be too coarse-grained to adequately represent a data practice. 

\noindent \textbf{Textual GUI components.} 
%
To detect and localize privacy-related texts from a given screenshot, we apply a state-of-the-art ultra-lightweight OCR system, named PaddlePaddle-OCR (PP-OCR)~\cite{li2022pp}. After obtaining these texts, we employ a pre-trained large language model, i.e., GPT-3.5\footnote{\url{https://platform.openai.com/docs/models/gpt-3-5}}, to classify their data types.
If the answer contains a data type, we will regard the \textit{\{detected text\}} as relevant to this data type; otherwise, it does not belong to any data type.

\noindent \textbf{Iconic GUI components - Localization.} 
An empirical study~\cite{chen2020object} demonstrated that rule-based object detection methods on GUIs outperform learning-based methods, such as YOLO~\cite{redmon2016you} series. In addition, rule-based methods have better explainability. To localize iconic GUI components from their screenshots, we adapt previous work~\cite{xie2020uied} and propose an updated set of detection rules to filter out irrelevant objects. Specifically, we have four rules presented as follows: (a) Filtering out elements whose area is greater than 10\% of the total area of the screenshot, in order to remove the bounding boxes and large images; (b) Filtering out elements whose area is less than 5\% of the total area of the screenshot, aiming to remove the small noises; (c) Removing rectangular elements whose aspect ratio (width to height) is less than 0.6, ruling out control bars and top/bottom banners; (d) Removing elements that area is overlapped with OCR results since some letters in large font-size are also detected as potential icons. 
By applying these rules to our dataset, we can obtain the position of the iconic GUI components on their screenshots.

\noindent \textbf{Iconic GUI components - Classification.}
%
CNN-based neural networks have a strong capability of recognizing patterns on GUIs~\cite{Liu:2018:LDS:3242587.3242650, xi2019deepintent, chen2022towards}. In this paper, we train two models to classify the detected graphical icons from the screenshots, including a ViT-based model and a ResNet-based model.
ViT stands for Vision Transformer ~\cite{dosovitskiy2020image}. 
Specifically, we employ a pre-trained vanilla ViT and fine-tune it on ImageNet~\cite{deng2009imagenet}. Regarding the ResNet-based model, we employ ResNet-18 as the backbone because of its tiny size and decent efficiency. We further add dropout layers to every ResNet block to enhance its generalizability, preventing the potential over-fitting issue. We then trained both models on the RICO-icon dataset with the default train-test split.
We use the ViT-based model as the default for Contextual Detection Model as it achieves the best performance.

\begin{table*}[t!]
\caption{Performance of SeePrivacy's modules. ``CDM'' stands for Context Detection Module. ``R-i Class'' stands for Rico-icon Class as per Table~\ref{tab_keyword_list}.}
\label{tab_evaluation_total}
\begin{minipage}{0.24\textwidth}
\subcaptionbox{\footnotesize CDM - Textual GUI components}{
\resizebox{0.95\textwidth}{!}{%
    \begin{tabular}{lccc}
        \toprule
        \textbf{Category} & \textbf{Accuracy} & \textbf{Precision} & \textbf{Recall} \\
        \midrule
        Name & 0.98 & 0.98 & 1.00 \\
        Birthday  & 1.00 & 1.00 & 1.00 \\
        Address  & 0.38 & 0.41 & 0.86 \\
        Phone  & 0.87 & 0.90 & 0.96 \\
        Email  & 0.61 & 0.97 & 0.62 \\
        Profile  & 0.72 & 0.77 & 0.92 \\
        Contacts  & 0.98 & 1.00 & 0.98 \\
        Location  & 0.92 & 0.96 & 0.96 \\
        Photos  & 0.87 & 1.00 & 0.87 \\
        Voices  & 0.91 & 0.91 & 1.00 \\
        Financial info  & 0.93 & 0.93 & 0.93 \\
        Social media  & 0.67 & 0.75 & 1.00 \\
        \midrule
        \textbf{Average} & 0.82 & 0.87 & 0.93 \\
        \bottomrule
    \end{tabular}
}%
}
\end{minipage}\hfill
\begin{minipage}{0.272\textwidth}
\subcaptionbox{CDM - Iconic GUI components}{
\resizebox{0.95\textwidth}{!}{%
\begin{tabular}{llrrr}
\toprule
\textbf{R-i Class} & \textbf{Category} & \textbf{Accuracy} & \textbf{Precision} & \textbf{Recall} \\
\midrule
Call  & Phone & 0.96 & 1.00  & 0.96 \\
Email  & Email & 0.96 & 0.95 & 1.00 \\
Avatar  & Profile & 0.92 & 0.96 & 0.96 \\
Follow & Contacts & 0.89 & 0.94 & 0.94 \\
Group  & Contacts &  0.73 & 0.73 & 1.00   \\
Location  & Location & 0.93 & 0.98 & 0.95 \\
Crosshair & Location & 0.91 & 0.94 & 0.97 \\
Photo & Photos & 0.92 & 0.92 & 1.00 \\
Wallpaper  & Photos & 0.94 & 0.94 & 1.00 \\
Videocam  & Photos &  0.83 & 1.00 & 0.83 \\
Microphone  & Voices & 1.00 & 1.00  & 1.00   \\
Cart &  Financial Info & 0.91 & 0.95   &   0.95   \\
Facebook  & Social media & 0.96 & 1.00  & 1.00  \\
Twitter  & Social media & 1.00  & 1.00 & 1.00 \\
\midrule
\textbf{Average} & &  0.92 & 0.95 & 0.96  \\
\bottomrule
\end{tabular}
}%
}
\end{minipage}\hfill
\begin{minipage}{0.24\textwidth}
\subcaptionbox{Segments Extraction Module}{
    \resizebox{0.95\textwidth}{!}{%
    \begin{tabular}{lccc}
    \toprule
    \textbf{Category}    & \textbf{Accuracy} & \textbf{Precision} & \textbf{Recall} \\
    \midrule
    Name        & 1.00 & 1.00 & 1.00  \\
    Birthday    & 0.96 & 1.00 & 0.95 \\
    Address     & 0.52 & 0.82 & 0.50 \\
    Phone       & 1.00 & 1.00 & 1.00  \\
    Email       & 0.99 & 0.99 & 1.00  \\
    Profile     & 0.99 & 0.99 & 1.00  \\
    Contacts    & 0.79 & 1.00 & 0.78  \\
    Location    & 0.94 & 0.95 & 0.98  \\
    Photos      & 0.95 & 0.97 & 0.97 \\
    Voices      & 0.86 & 0.76 & 1.00  \\
    Financial info   & 1.00 & 1.00 & 1.00 \\
    Social media& 1.00 & 1.00 & 1.00 \\
    \midrule
    \textbf{Average} & 0.94 & 0.98 & 0.96 \\
    \bottomrule
    \end{tabular}
    }%
}

\end{minipage}\hfill
\begin{minipage}{0.235\textwidth}
\subcaptionbox{\footnotesize Overall Context Identification}{
    \resizebox{0.95\textwidth}{!}{%
    \begin{tabular}{lccc}
    \toprule
    Category  & Accuracy  & Precision & Recall  \\
    \midrule
    Name        & 0.98        & 0.98     & 1.00       \\
    Birthday    & 1.00        & 1.00     & 1.00       \\
    Address     & 0.38        & 0.41     & 0.86       \\
    Phone       & 0.88        & 0.91     & 0.97       \\
    Email       & 0.66        & 0.96     & 0.68       \\
    Profile     & 0.70        & 0.74     & 0.93\\
    Contacts    & 0.75        & 0.94     & 0.78       \\
    Location    & 0.94        & 0.97     & 0.96       \\
    Photos      & 0.86        & 0.98     & 0.88       \\
    Voices  & 0.91       & 0.91     & 1.00       \\
    Financial info   & 0.81        & 0.93     & 0.87       \\
    Social media& 0.80        & 0.89     & 0.89       \\
    \midrule
    \textbf{Average} & 0.81 & 0.88 & 0.90 \\
    \bottomrule
    \end{tabular}
    }
}
\end{minipage}
\end{table*}

\subsection{Segment Extraction Module}
~\label{sec_framework_sem}
The objective of this module is to obtain privacy policy segments with their corresponding privacy-related contexts.
Mainstream mobile application markets, such as the Google Play app store~\cite{googledeveloper} and the Apple App Store~\cite{appleprivacy}, require developers to provide a privacy policy link on the application's homepage. This requirement allows us to access those links based on the name of a mobile application. To capture these privacy policies in HTML format, we leverage two Python libraries, \textit{Selenium}~\cite{Selenium} and \textit{BeautifulSoup}~\cite{beautifulsoup}, which fetch the information via the corresponding privacy policy links. Recognizing the global market trend and the need to cater to multilingual users, we note that an increasing number of privacy policies are available in multiple languages. To maintain consistency and clarity in our analysis, we deploy \textit{langdetect}~\cite{langdetect} library to filter out non-English texts in privacy policies. Note that our framework has the flexibility to be easily extended to accommodate other languages.

Previous studies~\cite{torre2020ai, xie2022scrutinizing} have discussed that only using sentence-level privacy policies processing and analysis can lead to contradictory conclusions, since the same or similar sentences in different sections may have different implications.
Therefore, to retrieve the relevant sentence-based segments, we adopt the multi-level privacy policies processing method proposed in~\cite{andow2019policylint, xie2022scrutinizing}.
For privacy policy documents whose HTML structure follows the $(\langle \texttt{Heading}\rangle \langle \texttt{Paragraph}\rangle ^ +)^+$ format, we reuse the pre-trained Bayesian multi-label classifier model~\cite{xie2022scrutinizing} to classify privacy policy paragraphs based on their headings since headings usually contain words that can be used as identifiers.  The classifier is claimed to achieve 0.85 accuracy and can classify the majority of the documents based on their headings. The paragraphs classified as \textit{Types} are related to the ``types of personal data collected by the current app''. Those paragraphs are highly related to specific data practices, so we only conduct sentence-level analysis on them.
For privacy policy documents whose HTML structure does not follow the $(\langle \texttt{Heading}\rangle \langle \texttt{Paragraph}\rangle ^ +)^+$ format, we follow previous works~\cite{harkous2018polisis, windl2022automating} to classify paragraphs on specific data practices. Specifically, we train a CNN-based multi-label classification model on a large privacy policy dataset~\cite{wilson2016creation} with 23K fine-grained data practice annotations. Note that this dataset contains 12 high-level categories for data practices, including \textit{First-Party Collection/Use} and \textit{Third-Party Sharing/Collection}. 
The results show that we can achieve a 0.84 top-1 precision for classifying the privacy segments with their related data practices. 

After obtaining paragraph-level privacy segments, we tokenize them into sentences using the \textit{stanza}~\cite{qi2020stanza} Python library. We then perform a two-stage sentence-level analysis to extract the sentences for each data type.
First, we perform a keyword search based on our collected keyword list (see Table~\ref{tab_keyword_list}).
If a keyword exists, the sentence will be added to the corresponding privacy policy segments of its data type.
Second, for sentences without any keywords, we further employ a Bayesian binary classifier to determine whether the sentence is related to a data type. The classifier is trained on the dataset from \cite{xie2022scrutinizing}, reaching an accuracy of 0.98.
If yes, we use \textit{SpaCy}~\cite{SpaCy} Python library to obtain noun chunks from the sentence. A noun chunk is a phrase that includes a noun and any connected words such as its adjectives.
We calculate the similarity of the phrases $phrase\_sim( \, )$ between noun chunks and keywords as follows:
\begin{equation*}
\label{formula_similarity}
\resizebox{.45\textwidth}{!}{%
    $\emph{\small phrase\_sim}\small (\emph{\small p}_1, \emph{\small p}_2\small) \small = \frac{2 \times \emph{path\_similarity}(\emph{p}_1, \emph{p}_2)}{\emph{word\_count}({\emph{p}_1}) + \emph{word\_count}(\emph{p}_2)}$
}%
\end{equation*}
where the $\emph{p}_1$ and $\emph{p}_2$ are a noun chunk and a keyword, respectively; $\emph{path\_similarity}(\,)$ is the function from WordNet~\cite{miller1995wordnet}, reflecting the similarity based on parsing their semantic constituency trees' structure; $\emph{word\_count}(\,)$ is to count the number of words.
If the similarity between a noun chunk and a keyword is higher than 0.8, an empirically set threshold, then we regard that the sentence that the noun chunk belongs to is the data type of the keyword.

Above all, we obtain the sentence-level policy segments related to each data type and group them as privacy policy segments. We also record the positions of keywords or noun chunks in the resulting sentences. For data types that do not have any match, we simply add "No relative information is found in the privacy policy." to notify users. 
For less popular applications, such notifications are likely to appear frequently, as they tend not to be stringent to disclosed privacy practices~\cite{zimmeck2019maps, liu2021have, USENIX_2022_GEODIFF}.

\subsection{CPPs Presentation Module}
For each mobile application, privacy policy, and its screenshots, the \textit{segment extraction module} and \textit{context detection module} allow us to obtain the specific privacy policy segments and associated privacy contexts. We further process this information to ensure a more coherent and user-friendly presentation.

Initially, we group the detected contexts of the same data types and map them to their corresponding privacy policy segments. As illustrated in Figure~\ref{fig_introduction}c, the \textsc{SeePrivacy} framework detects three contexts, including two textual GUI components, i.e., \textit{``Share your location''} and \textit{``use your birthday,''} and one iconic GUI component, such as a location mark icon in the middle right of the screenshot. 
These are classified into \textit{Location}, \textit{Location}, and \textit{Birthday} data types, respectively. 
The text \textit{``Share your location''} and the icon of the location mark are coupled with the segment about location information. The text \textit{``use your birthday''} only corresponds to the segment about \textit{Birthday} data type.
Drawing on existing research that emphasizes the improvement of readability through the strategic use of typography~\cite{palmen2023bold}, we render specific keywords or noun phrases within the privacy policy segments in bold font. This design choice, grounded in empirical findings, aims to accentuate essential information and thereby augment user comprehension.
\section{Evaluation}
\label{sec_evaluation}

In this section, we evaluate \textsc{SeePrivacy} both objectively and subjectively. Some visualization examples are displayed in Figure~\ref{fig_examples}.

\subsection{Quantitative Evaluation}
Our framework, consisting of three integral modules, is systematically evaluated for its performance on our specially curated benchmark dataset, named \textsc{Cpp4App}.
\subsubsection{Context Detection Module}
In assessing whether a detected GUI component aligns with the ground truth, we utilize Intersection over Union (IoU)~\cite{rezatofighi2019generalized}, a prevalent evaluation metric in object detection models. If the IoU between the ground truth and the detected component area exceeds a threshold $\beta$, the match is considered correct. Consistent with standard practice in object detection tasks~\cite{IoU_mrtrics}, we empirically establish $\beta = 0.5$.

\noindent \textbf{Textual GUI components.} 
Table~\ref{tab_evaluation_total}a presents the results related to the detection of textual GUI components. Our method exhibits a strong capability to recognize and categorize texts within mobile GUI screenshots, achieving an accuracy of 0.82, a precision of 0.87, and a recall of 0.93.

\noindent \textbf{Iconic GUI components.}
Table~\ref{tab_icon_deteciton_comparison} shows the comparison of the icon localization capability between our method and previous methods.
The results show that our proposed method performs better in identifying the icons curated on GUI screenshots.
Table~\ref{tab_evaluation_total}b shows the class-wise breakdown for Iconic GUI components, achieving an accuracy of 0.92, a precision of 0.95, and a recall of 0.96.
We then compare our icon classification models with several baselines on the RICO-icon dataset.
Table~\ref{tab_classification_comparison} shows that our ViT-based model achieves the best performance, but has the largest model size of 328 MB and the slowest time efficiency of 11.28 seconds.
The ResNet-based model provides a good balance of performance and efficiency, with the fast time efficiency, decent performance metrics, and a much smaller model size compared to others.


\begin{table}[t!]
\centering
\caption{Performance comparison between iconic GUI components localization methods.}
\label{tab_icon_deteciton_comparison}
\resizebox{0.85\linewidth}{!}{%
\begin{tabular}{lcc}
\toprule
\textbf{Method}   & \textbf{Precision (mislabel)} & \textbf{Recall (missing label)} \\
\midrule
UIED~\cite{xie2020uied}  &0.78 & 0.44\\
Liu et al.~\cite{Liu:2018:LDS:3242587.3242650}   & 0.79 &0.86 \\
Ours  &\textbf{0.84 }&\textbf{0.88}  \\
\bottomrule
\end{tabular}
}%
\end{table}



\begin{table}[t!]
\centering
\caption{Performance and efficiency comparison for iconic GUI elements classification on the RICO-icon dataset. ``kNN'' stands for k-nearest neighbors algorithm, and we use the normalized pixel values of the grayscale icon images as features. ``Acc.'', ``Prec.'' and ``Rec.'' represents Accuracy, Precision and Recall, respectively. Time(s) denotes the elapsed seconds to process a thousand icons.}
\label{tab_classification_comparison}
\resizebox{0.9\linewidth}{!}{%
\begin{tabular}{lccccc}
\toprule
\textbf{Model}                        & \textbf{Acc.} & \textbf{Prec.} & \textbf{Rec.} & \textbf{Size(MB)} &  \textbf{Time(s)} \\
\midrule
kNN~\cite{cover1967nearest} &0.75 &0.61 &0.52 &\textbf{0} & 6.78\\
Liu et al.~\cite{Liu:2018:LDS:3242587.3242650}   & 0.90       & \textbf{0.90}       & 0.79   & 293 &5.02 \\
MobileNetV2~\cite{sandler2018mobilenetv2} &0.86 &0.73 &0.73 & 9&0.56\\
MobileNetV3~\cite{howard2019searching} &0.79 &0.62 &0.66 &6 &0.52\\
\midrule
Ours (ViT-based) & \textbf{0.97} &\textbf{0.90} &\textbf{0.90} &328 &11.28\\
Ours (ResNet-based) &0.91  &0.80   & 0.81   & 43 &\textbf{0.37}\\
\bottomrule
\end{tabular}
}%
\end{table}

\begin{figure}[h!]%
    \centering
    \subfloat[\centering Screenshot 14-2 ]{{\includegraphics[width=0.15\textwidth]{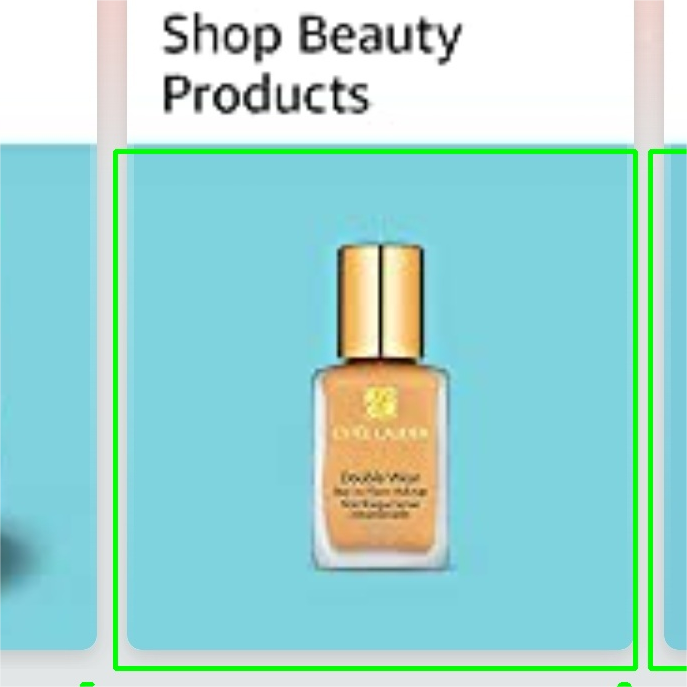}}}%
    \qquad
    \subfloat[\centering Screenshot 20-3]{{\includegraphics[width=0.15\textwidth]{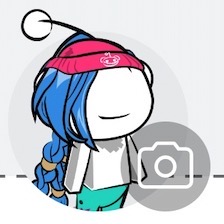}}}%
    \caption{Failure cases of the mislabel (left, the green bounding box) and the missing label (right, the camera icon).}%
    \label{fig_failure_example}%
\end{figure}

\noindent \textbf{Failure Case Analysis}
We manually examine failure cases in iconic GUI components detection and find that most of the missing labels are caused by overlapping with other components or interleaved boundaries. 
As for the mislabelled cases, small squarish images, e.g., item images in shopping apps, tend to be wrongly included as an icon. Figure~\ref{fig_failure_example} shows two examples of failure cases in our experiment. 

\noindent \textbf{Overall context identification.}
We further assess the overall performance of our context detection module by considering both textual and iconic GUI components. Table~\ref{tab_evaluation_total}d shows the overall context identification results considering textual and iconic GUI components. Specifically, \textsc{SeePrivacy} exhibits commendable results, achieving accuracy, precision, and recall rates of 0.81, 0.88, and 0.90, respectively. 

\subsubsection{Segment Extraction Module}
First, to determine whether the retrieval is successful, we introduce a similarity metric, denoted by $\emph{segment\_sim}(\,)$, which quantifies the similarity between the retrieved segments and the manually annotated ground truth. 
For a given ground truth $\emph{s}_{gt}$ and a retrieved segments $\emph{s}_{ret}$, our segment similarity $\emph{segment\_sim}(\,)$ is calculated as follows:
\begin{equation*}
~\label{formula_similarity}
\begin{aligned}
    \emph{segment\_sim}(\emph{s}_{ret}, \emph{s}_{gt}) &= \frac{1}{\textsc{Min}(n, m)}\sum_{i=1}^n\sum_{j=1}^m \overline {\emph{lcs}}(s^{ret}_{i}, s^{gt}_{j}) \\ 
    \overline {\emph{lcs}}(\emph{s}^{ret}_{i}, \emph{s}^{gt}_{j})  &=\frac{\emph{lcs}(s^{ret}_{i}, s^{gt}_{j})} {\textsc{Min}(\textsc{Len}(s^{ret}_{i}), \textsc{len}(s^{gt}_{j}))}
\end{aligned}
\end{equation*}
where $\emph{lcs}(\,)$ stands for the function to calculate longest common string; $\emph{s}^{ret}_{j}$ are phrases separated by punctuation of retrieved segments and $\emph{s}^{gt}_{i}$ are labelled phrases of ground truth segments. If it is greater than 0.8, an empirically set threshold, then the segment pair is regarded as a successful match.
Table~\ref{tab_evaluation_total}c shows the results of the segment extraction module, \textsc{SeePrivacy} achieves 0.94, 0.98, and 0.96 in terms of accuracy, precision, and recall, respectively.
There are some categories, such as \textit{Address}, which are more complex than the other categories due to relatively semantically ambiguous keywords. In addition, there are relatively less data for \textit{Voices} and \textit{Contacts}, thus the limited performance of this category could be caused by the randomness.

\subsection{Human Evaluation}
To further validate the functionality of \textsc{SeePrivacy}, we conduct a human evaluation to examine users' perceptions of its capacity for generating accurate CPPs for mobile applications.
This investigation seeks to provide empirical evidence on how \textsc{SeePrivacy} performs in real-world scenarios from users' perspective, thereby offering insights into its effectiveness and relevance.

\subsubsection{Evaluation Study Design}

\begin{table}[t!]
\centering
\caption{Statistic results of our human evaluation. SD stands for standard deviations.}
\label{tab_qualitative_results}
\resizebox{.8\linewidth}{!}{%
\begin{tabular}{lccc}
\toprule
\textbf{Topic} & \textbf{Mean}  & \textbf{Median}  & \textbf{SD}\\
\midrule
Data type \& Context&  4.22 & 5 & 1.14\\
Data type \& Policy segment& 3.95 & 4& 1.28\\
Policy segment \& Context& 3.75 & 4 & 1.40\\
\bottomrule
\end{tabular}
}%
\end{table}


For recruitment, we recruited examiners based on the following criteria: professional fluency in English reading and be capable of using mobile apps without accessibility tools (e.g., VoiceOver on iOS), via mailing lists of authors’ institution without offering monetary incentives.
For the human evaluation, we randomly sampled 30 screenshots with their privacy policies from the benchmark dataset and employ \textsc{SeePrivacy} to generate CPPs.
The manual examination involves heavy privacy policy reading and comprehending tasks, and we aim to reduce the session length to keep the examiners focused and increase the result validity. The average duration of evaluation session is about 30 minutes. 
To guarantee the validity of evaluation, we then divide the generated CPPs into three folds and compose the questionnaire for each fold, i.e., ten screenshots in each questionnaire.
For each CPP, examiners need to rate the following statements:
\begin{itemize}[leftmargin=*, noitemsep, topsep=0pt]
\itemsep0em
\item The data type matches with the detected context.
\item The data type matches with the displayed privacy policy segment.
\item The displayed privacy policy segment explains the detected context on the screenshot.
\end{itemize}
For all the ten allocated screenshots, they are asked to rate the agreement level on a 5-point Likert scale ranging from five (strongly agree) to one (strongly disagree).
In addition to providing a general introduction to the user study and obtaining consent, we also provide examiners with detailed descriptions of the data types, explicit instructions, and a concrete example to guide them prior to the beginning of our study.


\subsubsection{Evaluation Results}
In total, we recruited 15 people for the evaluation (6 females, 8 males, and 1 who preferred not to be mentioned). Each fold of screenshots was evaluated by five different examiners. The average duration for evaluation sessions is about 30 minutes.
The results are detailed in Table~\ref{tab_qualitative_results}. 
For the functionality evaluation pertaining to the matching between data types and detected contexts, the ratings were highly favorable, with more than 81\% of the responses falling within the categories of ``Agree'' or ``Strongly Agree.'' The mean and median values for this aspect were also substantial at 4.22 and 5, respectively. Concerning the alignment between data types and privacy policy segments, over 72\% of examiners assigned positive ratings, yielding a mean value of 3.95 and a median value of 4. Additionally, more than 77\% of examiners concurred that the retrieved privacy policy segments help explain the detected contexts, evidenced by a mean rating of 3.75 and a median rating of 4. Overall, these results demonstrate that \textsc{SeePrivacy} successfully fulfills its intended purpose by offering an efficient and effective solution for delivering CPPs to users.

%
\begin{figure} [t!]
\begin{subfigure}{.32\linewidth}
  \centering
  \includegraphics[width=0.9\linewidth]{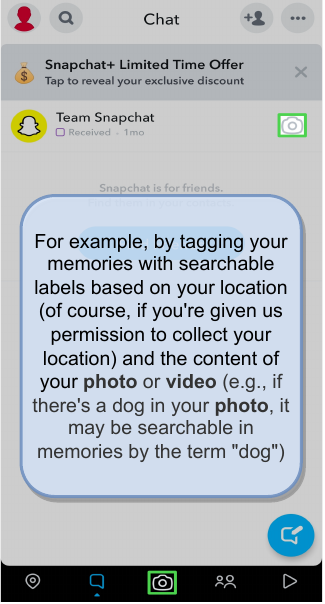}
  \caption{Sceenshot 6-1}
  \label{fig_example1}
\end{subfigure}%
\hspace{2pt}
\begin{subfigure}{.32\linewidth}
  \centering
 \includegraphics[width=0.9\linewidth]{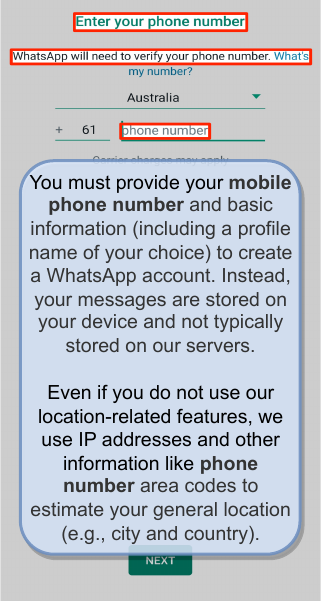}
  \caption{Screenshot 11-9}
  \label{fig_example2}
\end{subfigure}
\hfill
\begin{subfigure}{.32\linewidth}
  \centering
  \includegraphics[width=0.9\linewidth]{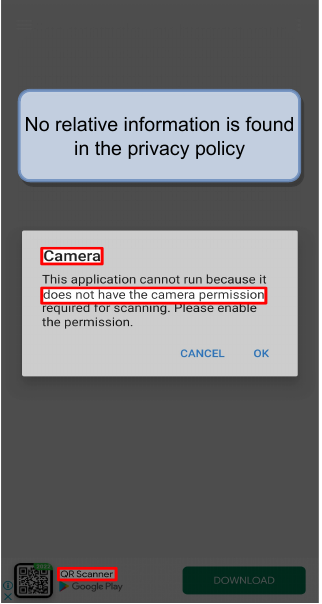}
  \caption{Screenshot 16-1}
  \label{fig_example3}
\end{subfigure}
\caption{
Three examples appeared in the human evaluation.
}
\label{fig_examples}
\end{figure}
%

\section{Discussion}~\label{sec_discussion}
In this section, we delve into the implications, potential applications, and contributions of our framework, elucidating its significance both for individual mobile application users and the wider community.

\subsection{Privacy Concern and Reading Willingness}
\begin{table*}[h]
\centering
\caption{Pre- and post-questions of the functionality evaluation about the general usability of SeePrivacy.}
\label{tab_questions}
\resizebox{.95\textwidth}{!}{%
\begin{tabular}{l|l|l|c|c|c}
\toprule
\textbf{No.} & \textbf{Pre- and post-questions} & \textbf{Scale (5-Likert)} & \textbf{Mean}  & \textbf{Median}  & \textbf{SD}\\
\midrule
\midrule
$\textit{Q}_1$  &\makecell[l]{How concerned are you about your privacy information \\ while using mobile apps?} & 5 for very concerned, 1 for very unconcerned & 4.13 & 4 &  1.09\\
\midrule
$\textit{Q}_2$    & \makecell[l]{Do you read mobile app's privacy policies when \\you encountered?} &  5 for always read, 1 for never read &2.00 & 2 & 1.03 \\
\midrule
\midrule
$\textit{Q}_3$   & \makecell[l]{What do you think the usefulness of this tool in terms of \\ providing privacy information for mobile apps?} & 5 for very useful, 1 for very useless & 4.53 & 5 & 0.62  \\
\midrule
$\textit{Q}_4$  & \makecell[l]{Will you read contextual privacy policies when you \\encountered in future?} & 5 for always read, 1 for never read & 4.07 & 4 & 0.85  \\
\bottomrule
\end{tabular}
}%
\end{table*}

Before and after the human evaluation, we also asked several questions about privacy concerns and their privacy policy reading habits to examiners. 
Table~\ref{tab_questions} lists the specific questions ($\textit{Q}_1$ and $\textit{Q}_2$ are pre-evaluation; $\textit{Q}_3$ and $\textit{Q}_4$ are post-evaluation) and results.
It is evident that privacy concerns are a prominent issue for users while using mobile applications, with a mean concern rating of 4.13 and a median value of 4 for question $\textit{Q}_1$. 
The utility of \textsc{SeePrivacy} in furnishing relevant privacy information for mobile applications was highly acclaimed by the participants. Specifically, 14 out of 15 participants evaluated it as either ``Very useful'' or ``Useful.'' 

In addition, we compare the reading willingness between traditional privacy policies and the CPPs generated by \textsc{SeePrivacy}. The results demonstrate a substantial increase in the willingness of participants to engage with CPPs, as reflected by a mean score that surged from 2 to 4.1. 
Furthermore, among 8 participants who are ``very concerned'' about their privacy information while using mobile apps, 6 select that they will always read CPPs, and 2 select often read CPPs.
These findings collectively underscore the potential of \textsc{SeePrivacy}. By enhancing the transparency, accessibility, and digestibility of privacy policies, our proposed framework plays a vital role in engaging users to read privacy statements, potentially mitigating the prevalence of ``digital resignation''.
Although the diversity of participants is slightly under representative, we believe those preliminary findings are still valuable to be reported in this paper. 

%
\begin{figure}[t!]
  \centering
  \includegraphics[width=1.0\linewidth]{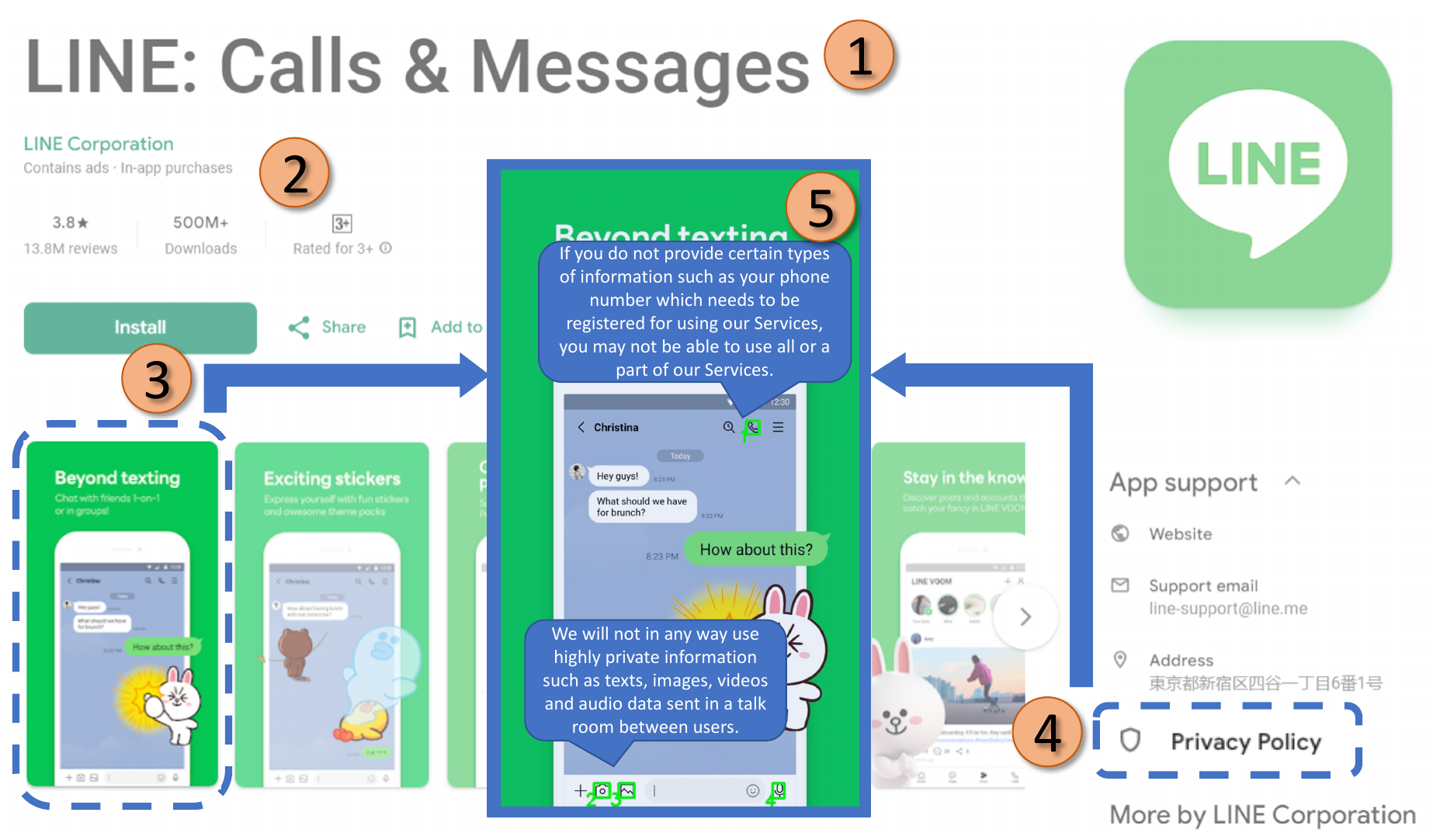}
  \caption{An example of the application scenario of our framework. (1) The homepage of the app at the Google Play app store. (2) Basic information of the app. (3) The showcase screenshots provided by the app developer. (4) The link of privacy policy. (5) The CPPs generated by \textsc{SeePrivacy}.}
  \label{fig_cpp_application}
\end{figure}
%
\subsection{Potential Adoption Scenario}
\subsubsection {CPP in App Market}

Our framework has the potential to be applied in various real-world scenarios. 
One plug-and-play application is displaying CPP in the app market, as illustrated in Figure~\ref{fig_cpp_application}.
Research has shown that users' app downloading decisions are often influenced by curated market information such as ratings and reviews~\cite{genc2017systematic}.
On the apps' homepage (\circled{1}, \circled{2}) in app markets (e.g., the Google Play app store), developers usually curate some showcase screenshots (\circled{3}) of their apps to exhibit some characteristics. Those screenshots often include the actual GUIs that appear during the users' actual usage. As a privacy policy link (\circled{4}) is also commonly mandated by app markets, our framework will be able to generate the CPPs (\circled{5}) by utilizing the showcase screenshots and the corresponding privacy policy on the homepage. 
By leveraging \textsc{SeePrivacy}, users can gain contextualized insights into potential data practices and pertinent policy statements of the mobile application prior to installation. This application allows users to comprehend data practices with real-use scenarios, enabling them to make informative decisions and establish better expectations before downloading the app.


\subsubsection{Lack of Disclosure: Contextual Detection}

\begin{figure}[tbp]
    \centering
    \begin{subfigure}[b]{0.23\textwidth}
        \centering
        \includegraphics[width=\textwidth]{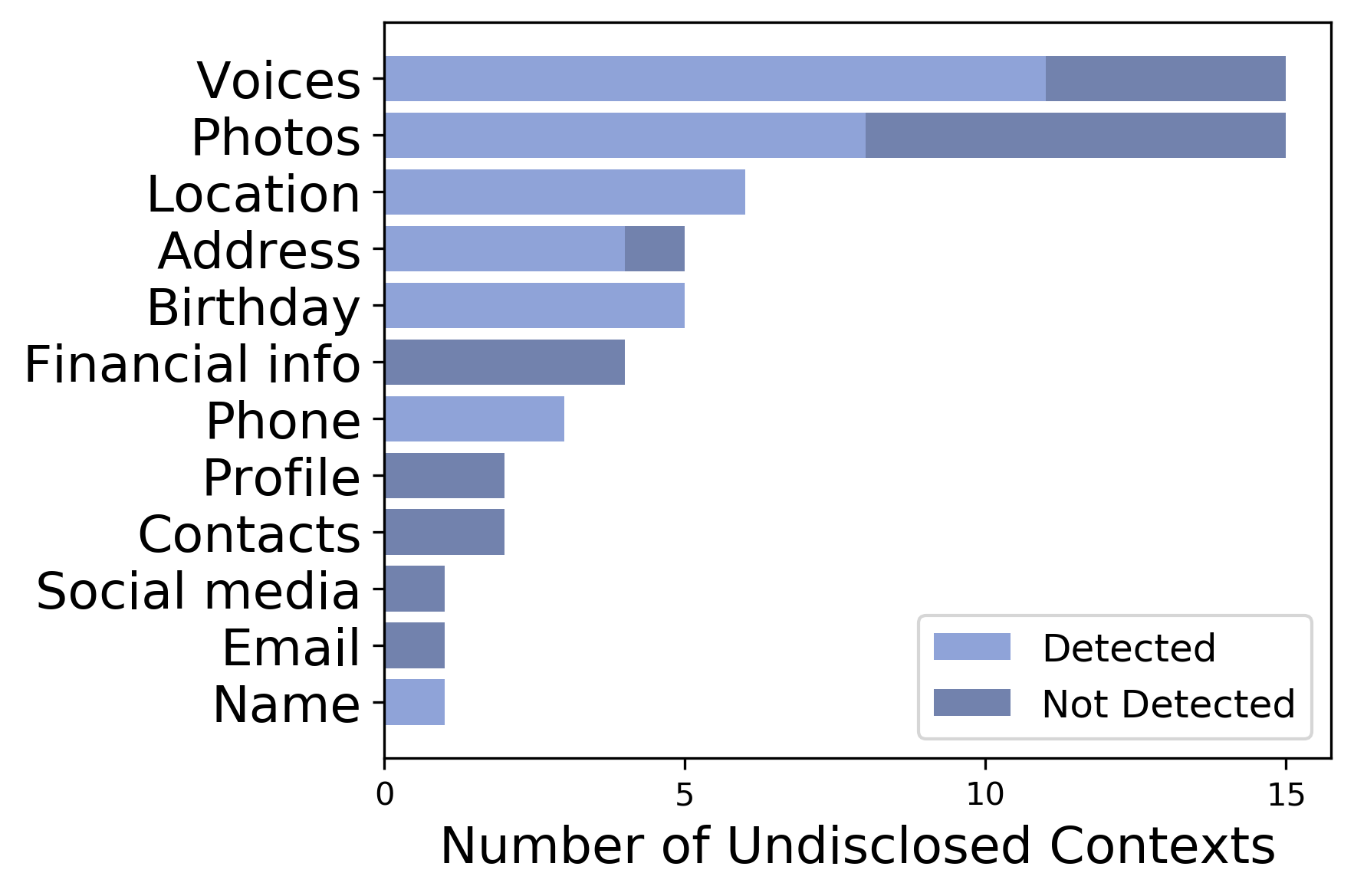}
        \caption{Detection Result of Contexts}
        \label{fig:detect_lack}
    \end{subfigure}
    \hfill
    \begin{subfigure}[b]{0.24\textwidth}
        \centering
        \includegraphics[width=\textwidth]{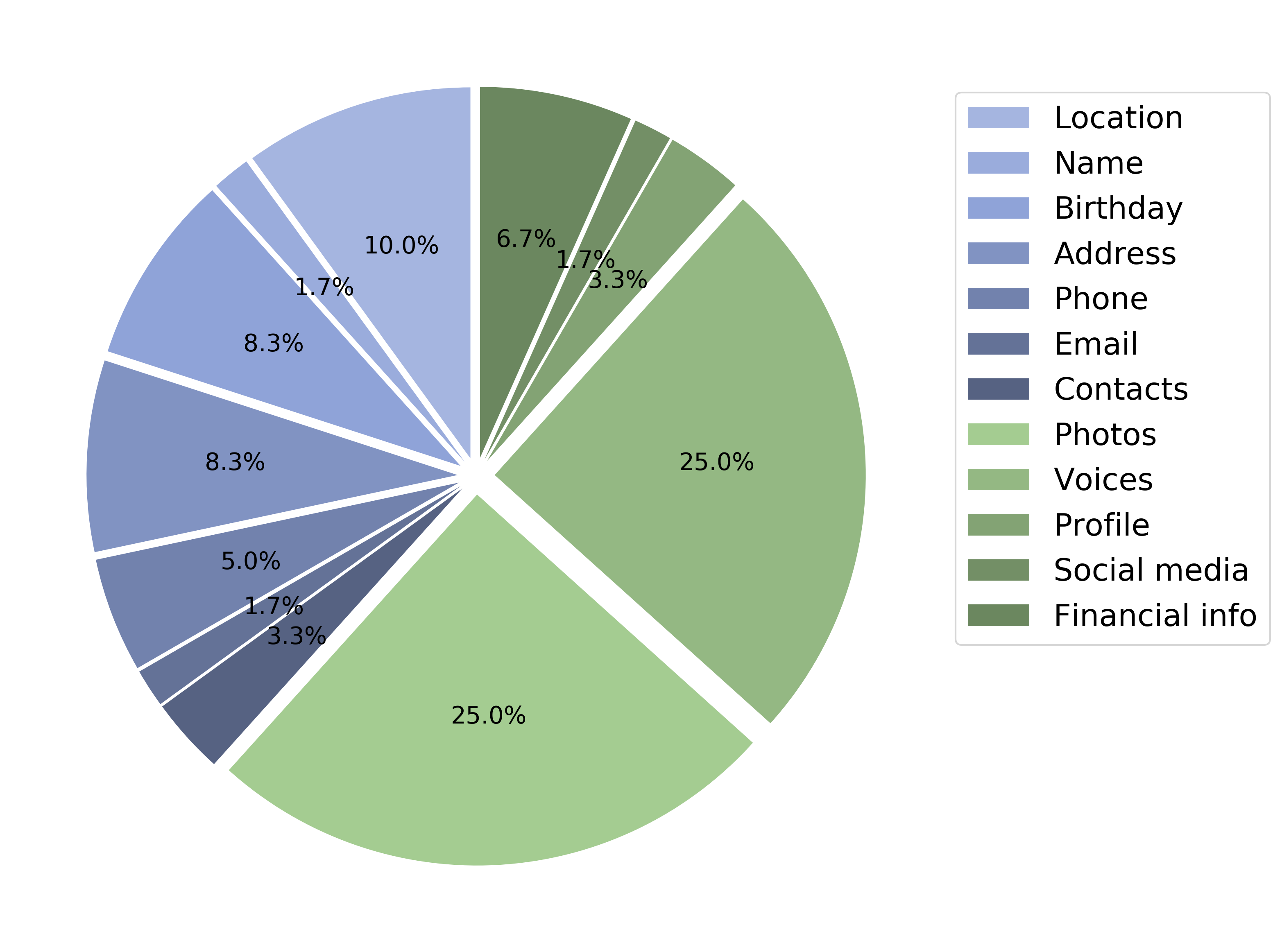}
        \caption{Distribution of Contexts}
        \label{fig:dist_lack}
    \end{subfigure}
    \caption{The left figure shows the detection result of contexts that do not have information in privacy policies. The right figure presents the data type distribution of those contexts.
    }
    \label{fig:bothfigures_lack}
\end{figure}

\label{sec_lack_disclosure}

Numerous prior studies~\cite{zimmeck2019maps, zhang2020does, xie2022scrutinizing, xiaolalaine} have highlighted the inconsistency between statements articulated in privacy policies and an application's actual behaviors. Specifically, the absence of proper disclosure concerning data practices may place users at considerable risk concerning digital security. In the current landscape of privacy policy analysis, we identify two critical shortcomings.
First, in the case of essential applications, users often find themselves in a position where they must accept the terms, even if they are made aware of any lack of disclosure within the policies. There is no space to negotiate with developers for specific functions or certain GUI components under the ``all-or-nothing'' model.
Second, existing methodologies fail to link the noted inconsistencies to the real contexts with which users may engage on GUIs. This omission presents a significant gap in understanding how policy misalignment may manifest in a user's interaction with an application.
The lack of notice and choice inevitably leads to a digital resignation phenomenon, where users acquiesce to potentially problematic terms out of necessity, rather than informed consent.

To address the aforementioned challenges, \textsc{SeePrivacy} is proposed to provide the affected GUI elements, i.e., the context, allowing users to make informed decisions about their interactions. For example, in Figure~\ref{fig_example3}, \textsc{SeePrivacy} detects three contexts related to \textit{``Photos''} in red bounding boxes, but does not obtain any relevant statements in the privacy policy of the app. Therefore, ``No relative information is found in the privacy policy'' is displayed to warn users about the potential risk.
Users can simply choose not to interact with those \textit{``Photos''}-related GUI elements as a band-aid solution, meanwhile ``safely'' using other features of the app. In total, we found that 60 contexts do not have corresponding privacy policy segments in \textsc{Cpp4App}.
The 60 contexts are distributed among 17 mobile applications in \textsc{Cpp4App} and cover all 12 data types. Figure \ref{fig:bothfigures_lack} illustrates the specific numbers and the distribution for lack of disclosure. Among the various data types, \textit{Photos} and \textit{Voices} combined constitute 50\% of the data. Interestingly, the \textit{Voices} data type seems to be frequently overlooked. Contexts for this data type typically involve requests to record audio (e.g., a microphone icon), or the conversion of voice to text for search functionality (e.g., a phone icon).
In total, \textsc{SeePrivacy} successfully identified 63.3\% (38/60) lack of disclosure contexts.

\subsection{Broader Impacts}

\noindent \textbf{Legal Impacts.} 
Privacy regulations frequently stipulate specific requirements regarding the completeness of privacy practices within policies. For instance, GDPR mandates that privacy notices disclose the categories of personal data concerned [GDPR, Art. 14(1)(d)] and recipients or categories of recipients of the personal data [Art. 13(1)(e), 14(1)(e)]. Likewise, the CCPA prescribes that privacy notices should enumerate the categories of personal information collected in the preceding 12 months [CCPA §1798.130(a)(5)(B), §1798.110(c), Regs §999.308(c)(1)(d)].
While such requirements are often adhered to by mobile application developers and can be directly assessed by regulators, \textbf{the challenges of formulating best practices to achieve transparency, and evaluating the transparency level of a privacy statement, persist}. 
In this context, our innovative definition of CPPs within the mobile scenario emerges as a potential metric for evaluating transparency. By linking the stated privacy practices to the actual GUI elements with which users interact, \textsc{SeePrivacy} offers an empirical means of ensuring that privacy statements are not only compliant with legal requirements but also meaningfully transparent to end-users. This approach aligns with the broader legal and ethical imperatives of user-centric privacy and could contribute significantly to the development of transparent and accountable privacy practices in the mobile application ecosystem.

\noindent \textbf{Accessibility and Inclusion.} 
By transforming privacy policies into a contextual and understandable format, our approach potentially broadens their accessibility to a more diverse range of users. This includes individuals who may not possess advanced literacy skills as well as those who face challenges related to reading disorders. Such inclusivity not only democratizes the understanding of privacy policies but also aligns with broader societal goals of ensuring that vital information is comprehensible to a wide audience. In this regard, \textsc{SeePrivacy} contributes to creating a more equitable digital environment where awareness and control over personal data are not confined to a specific subset of users but are extended to encompass various demographic groups. The potential reach and impact of this approach underscore the need for continued research and innovation in the domain of user-centered privacy communication.

\noindent \textbf{Children and Education.} 
The GDPR Recital 58 (The Principle of Transparency) states ``\textit{... any information and communication, where processing is addressed to a child, should be in such a clear and plain language that the child can easily understand.}'' 
This principle acknowledges the unique challenge of making legal and privacy-related texts accessible to young audiences.
The \textsc{SeePrivacy} framework aligns with this directive by transforming privacy policies into potentially engaging and comprehensible formats to children.
By connecting formal legal terms with real privacy use cases, it can facilitate a more straightforward understanding of the implications of online activities. 
While \textsc{SeePrivacy} strives to comply with regulatory requirements, its broader goal is to contribute to the digital literacy of disadvantage groups. 
This effort seeks to provide them with a basic understanding of online privacy, aiming to equip the next generation with the knowledge needed to approach the digital world with informed caution and responsibility. 
\textsc{SeePrivacy} aligns with the growing emphasis on the creation of tools and resources specifically designed to meet the developmental needs and competencies of young internet users.

\subsection{\textit{Notice-and-Choice Privacy Framework} for Mobile Apps: Is This the Way?}

The \textit{notice-and-choice} framework for mobile application ecosystem is fundamentally challenged by two critical issues.
First, there is a significant disconnection between privacy notices and user-oriented interactions, leading to ineffective engagement of users in meaningful ``notice'' processes and subsequent privacy decision-making.
Second, as illustrated in Figure~\ref{fig_intro_install} and~\ref{fig_intro_invoke}, 
the current designs of privacy controls often present users with an all-or-nothing choice, leaving them without genuine control over their data privacy. Therefore, this imbalance between utility and privacy highlights the need for administrative action through the regulation establishment and enforcement to safeguard consumer privacy rights.

While \textsc{SeePrivacy} is not a “panacea” for these challenges, it enlightens ``a new hope'' in this landscape. 
It suggests a paradigm shift in the privacy \textit{notice} (what users are agreeing to) and \textit{choice} (to agree or not) from a generic application level to specific functions or certain GUI components.
It is critical to proactively elevate privacy considerations from non-functional to functional requirements in software design and development.
Our framework aims to actively prompt users with timely and contextually relevant privacy information, enabling the potential for more granular control in future.
In addition, exploring the practice of Right to be Forgotten can sufficiently supplement this approach, empowering users with more control over their choices~\cite{zhang2023right, zhang2024forgotten}.

\subsection{Limitations.}
Our current human evaluation focuses more on the functionality of \textsc{SeePrivacy}, i.e., the level of matching between contexts and retrieved policy segments, thus, all examiners are relatively well-educated in technology.
Thus, the preliminary findings of reading willingness and popularity of CPPs among them may be biased due to the small and non-representative sample.
We plan to further conduct a large-scale user study that includes people from diverse backgrounds in the future.
In addition, the exploration of fatigue and habituation within human-computer interaction and its impact on interaction design have not been fully developed. These factors may profoundly affect user engagement and interface efficacy.
Furthermore, the time and storage efficiency of \textsc{SeePrivacy} were evaluated theoretically and may not reflect real-world performance in different application scenarios.
Regarding generalizability, our CPP detection is entirely vision-based, allowing for straightforward adaptation to other platforms, such as iOS mobile applications.
\section{Conclusion and Future Work}
~\label{sec_conclusion}
Privacy policies have long been criticized for their complexity and poor readability, leading to a growing interest in contextual privacy policies (CPPs) as user-focused alternatives. 
In this paper, we introduce the concept of CPPs specifically within the domain of mobile applications and propose a novel framework, \textsc{SeePrivacy}, designed to automatically generate these privacy policies. We rigorously evaluated \textsc{SeePrivacy} on our benchmark dataset, \textsc{Cpp4App}, which consists of over 1,200 privacy-related contexts and corresponding privacy policy segments. 
In the quantitative evaluation, our framework demonstrates 0.88 precision and 0.90 recall to detect contexts; as well as 0.98 precision and 0.96 recall to extract corresponding policy segments.
Furthermore, a human evaluation was conducted to validate the functionality our approach, reflected by a median score of 4 on a 5-point Likert scale.
Our pre- and post-survey revealed that people exhibit much greater willingness to engage with CPPs, scoring 4.1 out of 5, compared to a mere 2 out of 5 for traditional privacy policies.
We also discussed the potential adoption scenarios and implications of \textsc{SeePrivacy} to the wider community. 

Our promising results identify two potential research directions.
First, while our current efforts emphasize methodological development and evaluation, we recognize the importance of exploring real-world deployment scenarios. 
The user experience, particularly how users interact with contextual privacy policies, is vital and could significantly impact the system's effectiveness. 
Future work may include integrating \textsc{SeePrivacy} into specific human-device interactions, such as live applications on smartphones or sandbox environments within which applications operate. 
Second, although we aim to deliver the original privacy policy content without any modification, we notice that some retrieved policy segments are too long to comprehend in a timely manner. Thereby, it is worthwhile to evaluate how to digest lengthy policy segments with minimum information loss and distortion to maintain their integrity and practicability simultaneously.
Overall, we believe this work provides a substantial step forward in delivering transparent privacy notices, paving the way for future research and practical implementation in this crucial area.

\noindent \textbf{Data Availability.} Benchmark dataset \textsc{Cpp4App} and the replication package are available at: \url{https://github.com/Cpp4App/Cpp4App}.



\section*{Acknowledgments}

This research / project is supported by CSIRO's Data61 PhD Scholarship, and the National Research Foundation, Singapore, and Cyber Security Agency of Singapore under its National Cybersecurity Research and Development Programme, NCRP25-P03-NCR-TAU. Any opinions, findings and conclusions or recommendations expressed in this material are those of the author(s) and do not reflect the views of CSIRO's Data61, National Research Foundation, Singapore and Cyber Security Agency of Singapore.


\bibliographystyle{plainnat}
\bibliography{11_References}

\section{Appendix}

\subsection{Data Types in Privacy Regulations}~\label{sec_appendix_regulation}

\noindent \textbf{What is ``personal identifier'' in this study?}
Personal Identifier (PID) commonly appears in data protection and privacy legislation. They have different definitions and may be further specified according to the circumstances~\cite{us_personalinfo, gdpr_personalinfo, uop_personalinfo, au_personalinfo}.
If there is no further explanation in the legislative context, we only take the less controversial subset of common PID examples, \textit{Name}, \textit{Date of Birthday}, \textit{Address}, \textit{Phone Number}, \textit{Email}, and unique \textit{Account Info} to avoid the unintended comprehension.

The sources that we refer for the definition and interpretation of ``personal information'' in privacy regulations are listed as below:

\begin{itemize}[leftmargin=*]
\item \textbf{GDPR:}~\href{https://gdpr.eu/eu-gdpr-personal-data}{gdpr.eu/eu-gdpr-personal-data}
\item \textbf{CCPA:}~\href{https://oag.ca.gov/privacy/ccpa}{oag.ca.gov/privacy/ccpa}
\item \textbf{CalOPPA:}\href{https://leginfo.legislature.ca.gov/faces/codes_displayText.xhtml?division=8.&chapter=22.&lawCode=BPC}{leginfo.legislature.ca.gov/faces/codes}
\textbf{ and }
\href{https://termageddon.com/caloppa-personal-information/}{termageddon.com/caloppa-personal-information/}
\item \textbf{COPPA:}\href{https://www.ftc.gov/business-guidance/resources/complying-coppa-frequently-asked-questions}{ftc.gov/business-guidance/resources/complying-coppa-frequently-asked-questions}
\item \textbf{APP:}\href{https://www.oaic.gov.au/privacy/your-privacy-rights/your-personal-information/what-is-personal-information}{oaic.gov.au/privacy/your-privacy-rights/your-personal-information/what-is-personal-information}
\end{itemize}






\vspace{20pt}
\subsection{Human Evaluation Breakdown}~\label{sec_appendix_user}

Table~\ref{tab_appendix_user_study} shows the statistic results of the agreement of retrieved privacy policy segments explain detected contexts for each fold.


\begin{table}[h]
\centering
\caption{Statistic results of our user study in each fold.}
\label{tab_appendix_user_study}
\resizebox{\linewidth}{!}{%
\begin{tabular}{lcc|cc|cc}
\toprule
& \multicolumn{2}{c}{\textbf{Fold 1}} & \multicolumn{2}{c}{\textbf{Fold 2}} & \multicolumn{2}{c}{\textbf{Fold 3}}\\
\cmidrule(lr){2-3} \cmidrule(lr){4-5} \cmidrule(lr){6-7}
\textbf{Topic} & \textbf{Mean} & \textbf{Median} & \textbf{Mean} & \textbf{Median} & \textbf{Mean} & \textbf{Median}\\
\midrule
Data type \& Context & 4.16 & 5 & 4.43 & 5 & 4.09 & 4 \\
Data type \& Policy segment & 3.87 & 4 & 4.04 & 5 & 3.96 & 4\\
Policy segment \& Context & 3.73 & 4 & 3.63 & 4 & 3.80 & 4\\
\bottomrule
\end{tabular}
}
\end{table}

\subsection{Visualization Examples of Contextual Detection Module}

\begin{figure*} [t!]
\centering

\begin{subfigure}{.2\linewidth}
  \centering
  \includegraphics[width=.98\linewidth]{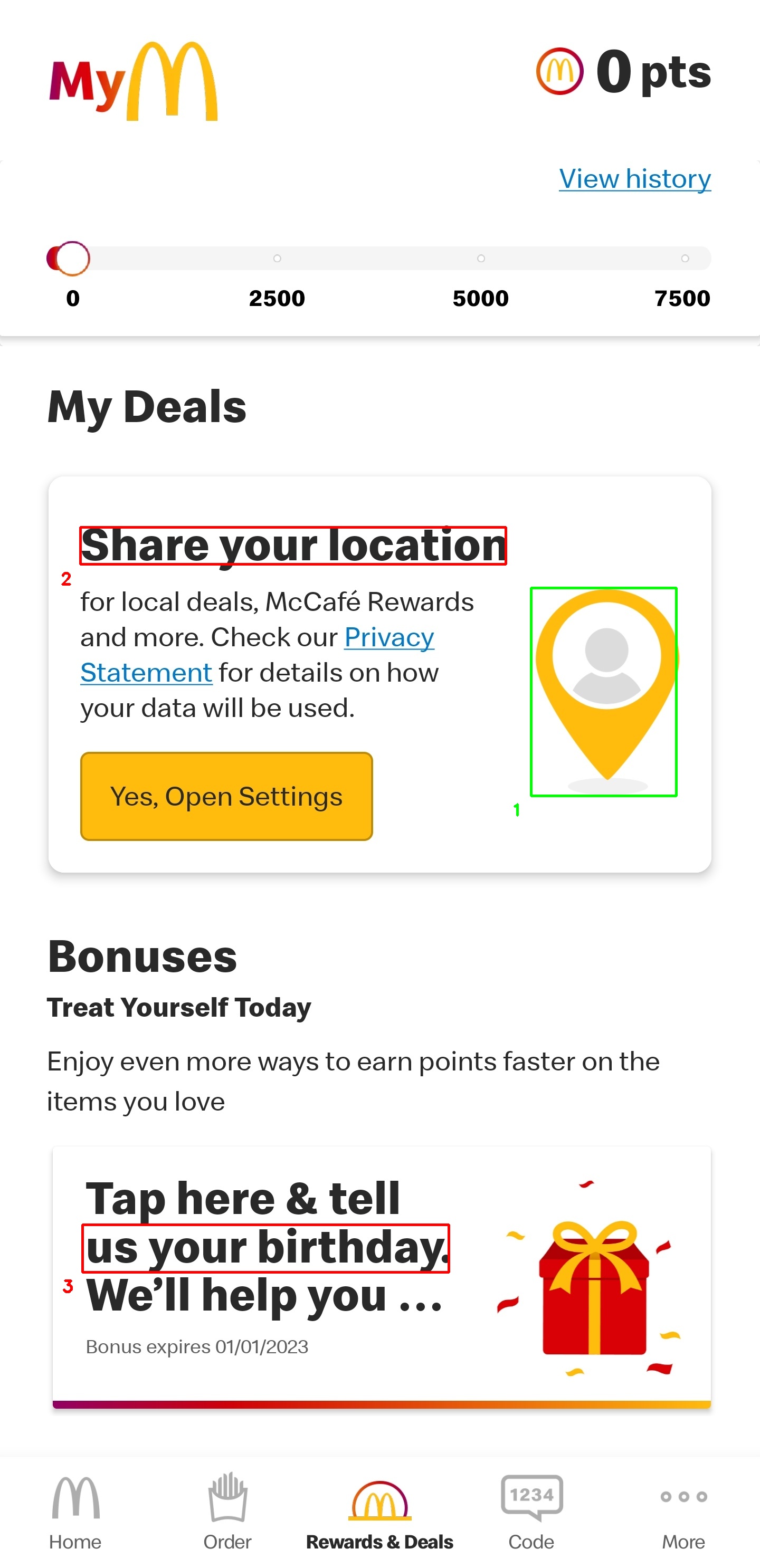}
  \caption{Screenshot 1-1}
  \label{example_1}
\end{subfigure}%
\hspace{.1\linewidth}
\begin{subfigure}{.2\linewidth}
  \centering
  \includegraphics[width=.98\linewidth]{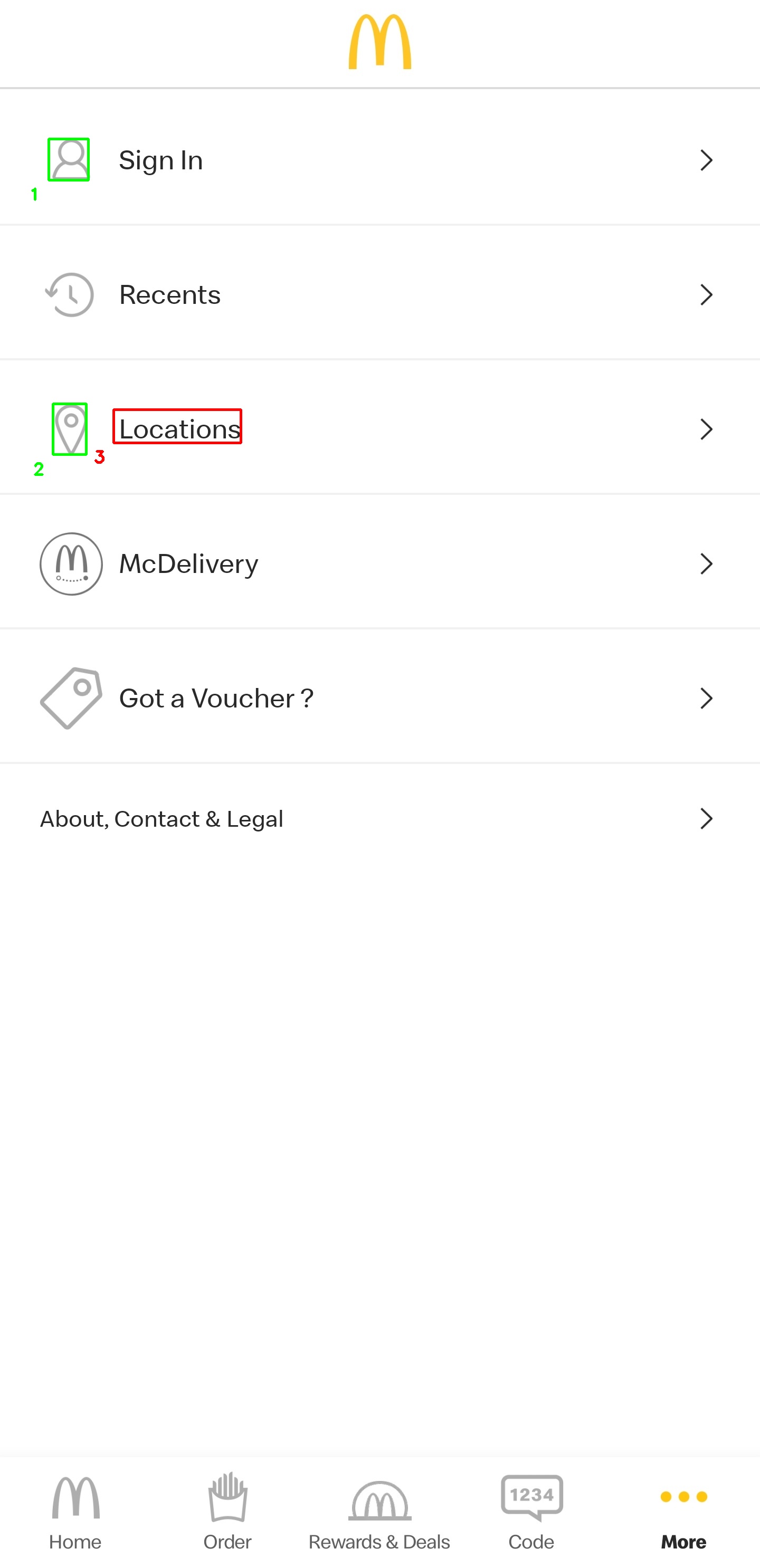}
  \caption{Screenshot 1-4}
  \label{example_2}
\end{subfigure}
\hspace{.1\linewidth}
\begin{subfigure}{.2\linewidth}
  \centering
  \includegraphics[width=.98\linewidth]{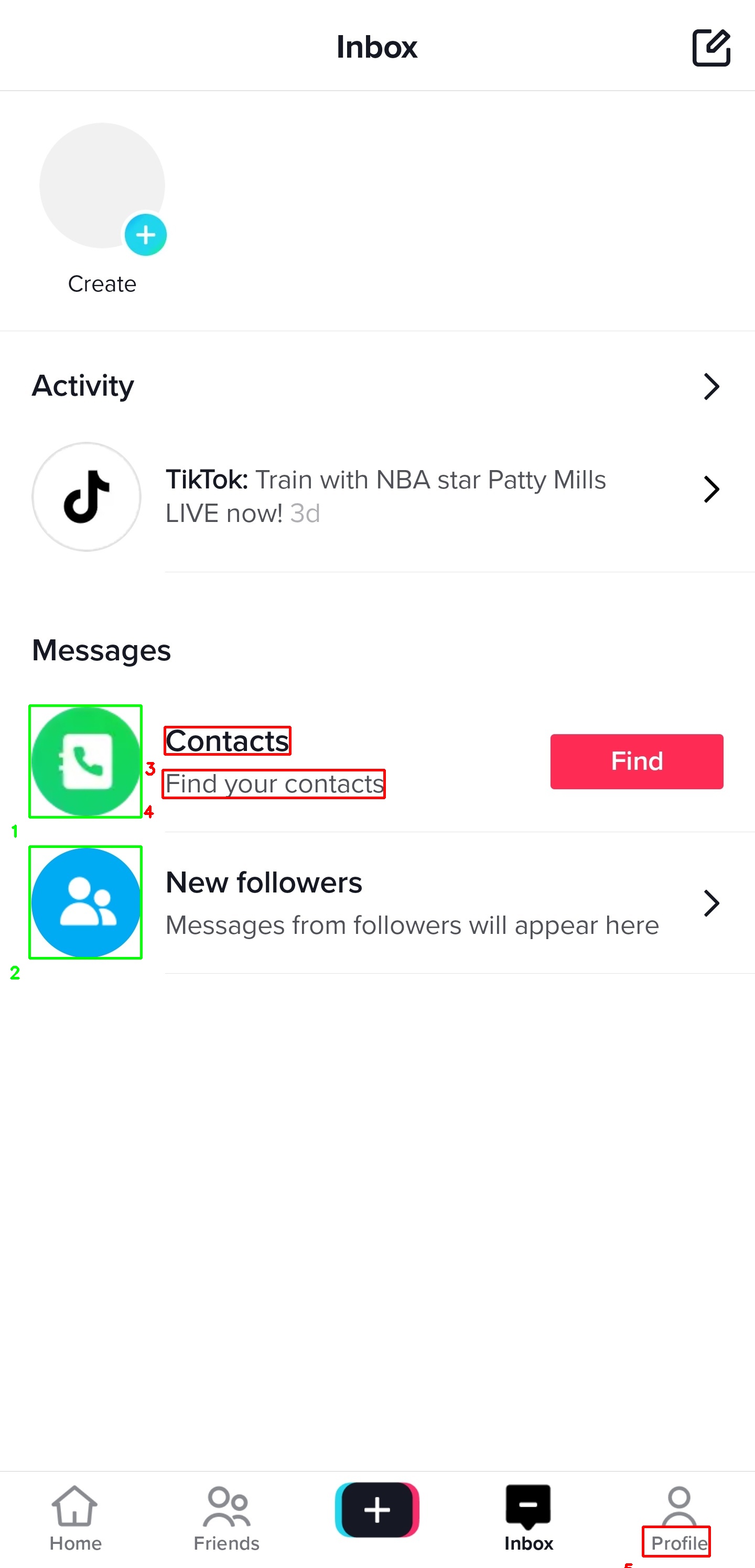}
  \caption{Screenshot 4-2}
  \label{example_3}
\end{subfigure}

\bigskip

\begin{subfigure}{.2\linewidth}
  \centering
  \includegraphics[width=.98\linewidth]{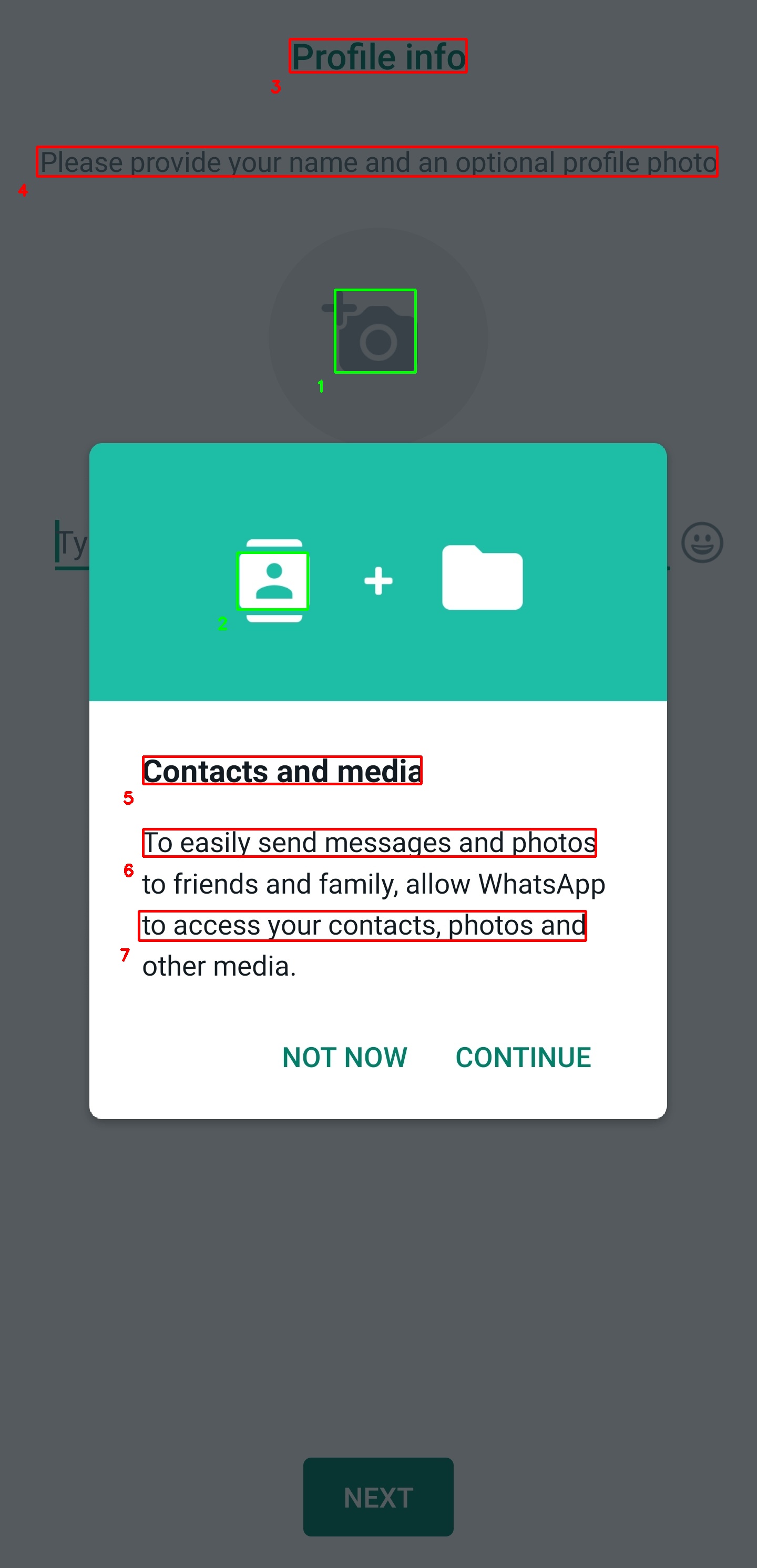}
  \caption{Screenshot 6-2}
  \label{example_4}
\end{subfigure}%
\hspace{.1\linewidth}
\begin{subfigure}{.2\linewidth}
  \centering
  \includegraphics[width=.98\linewidth]{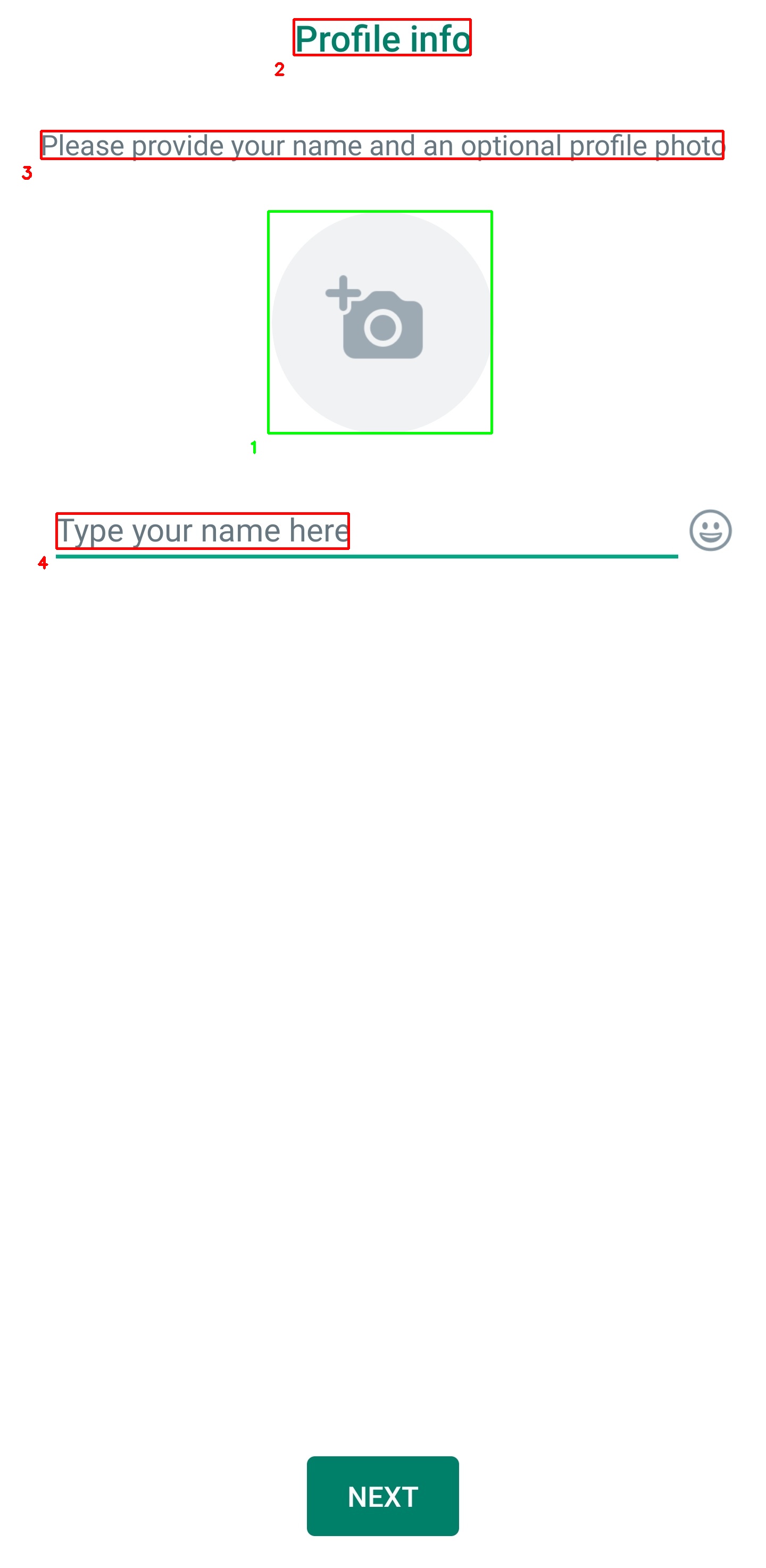}
  \caption{Screenshot 6-3}
  \label{example_5}
\end{subfigure}
\hspace{.1\linewidth}
\begin{subfigure}{.2\linewidth}
  \centering
  \includegraphics[width=.98\linewidth]{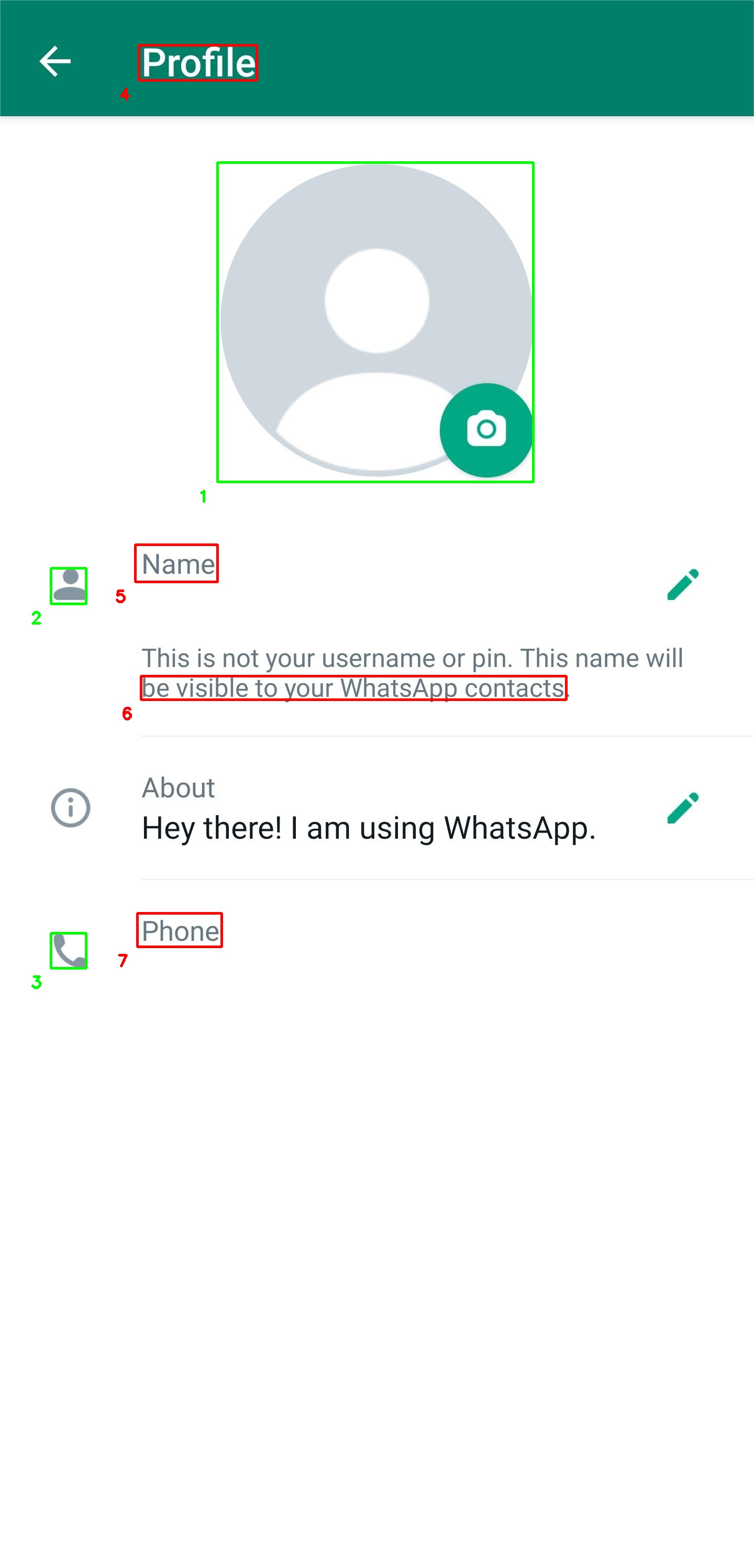}
  \caption{Screenshot 6-10}
  \label{example_6}
\end{subfigure}
\caption{Visualization examples: The screenshot results after CDM.}
\label{fig_appendix_example_2}
\end{figure*}

A collection of high-resolution examples of contexts detected by
our framework is shown in Figure~\ref{fig_appendix_example_2}.




%
%

\end{document}